\documentclass[11pt,amsmath,amssymb]{revtex4}
\usepackage{amsfonts}
\usepackage{amsmath}
\usepackage{amssymb}
\usepackage{array}
\usepackage{bm}
\usepackage{braket}
\usepackage{breqn}
\usepackage{color}
\usepackage{dcolumn}
\usepackage{diagbox}
\usepackage{epsfig}
\usepackage{epsf}
\usepackage{epstopdf}
\usepackage{float}
\usepackage{graphicx}
\usepackage{graphics}
\usepackage{harpoon}
\usepackage{indentfirst}
\usepackage{makecell}
\usepackage{mathrsfs}
\usepackage{mathtools}
\usepackage{multirow}
\usepackage{pifont}
\usepackage{xcolor}
\usepackage{hyperref}

\newcommand{\RM}[1]{\textrm{\uppercase\expandafter{\romannumeral#1}}}

\begin{document}

\title{In-depth Study of Spin- and Momentum-Dependent Interaction Potentials Between Two Spin-1/2 Fermions Mediated by Light Spin-0 Particles}

\author{
Yang Zhong$^{1}$,
Zhi-Hui Guo$^{2}$\protect\footnotemark[1]\protect\footnotetext[1]{E-mail: zhguo@hebtu.edu.cn},
Hai-Qing Zhou$^{1}$\protect\footnotemark[1]\protect\footnotetext[1]{E-mail: zhouhq@seu.edu.cn} \\
$^1$ School of Physics, Southeast University, NanJing 211189, China\\
$^2$ Department of Physics and Hebei Key Laboratory of Photophysics Research and Application, Hebei Normal University, Shijiazhuang 050024, China
}

\date{\today}

\begin{abstract}
We present a calculation by including the relativistic and off-shell contributions to the interaction potentials between two spin-1/2 fermions mediated by the exchange of light spin-0 particles, in both momentum and coordinate spaces. Our calculation is based on the four-point Green function rather than the scattering amplitude. Among the sixteen potential components, eight that vanish in the non-relativistic limit are shown to acquire nonzero relativistic and off-shell corrections. In addition to providing relativistic and off-shell corrections to the operator basis commonly used in the literature, we introduce an alternative operator basis that facilitates the derivation of interaction potentials in coordinate space. Furthermore, we calculate both the long-range and short-range components of the potentials, which can be useful for future experimental analyses at both macroscopic and atomic scales.
\end{abstract}

\maketitle

\section{Introduction}

In recent years, tabletop-scale laboratory experiments have emerged as a promising approach for exploring new physics phenomena beyond the Standard Model (BSM). These experiments are typically sensitive to long-range forces mediated by light BSM particles, such as axions, dilatons, and dark photons, etc~\cite{Safronova:2017xyt}. Notably, precision measurements at atomic and macroscopic scales have begun to probe exotic spin- and velocity-dependent interaction potentials with high sensitivity~\cite{Dobrescu:2006au,Raffelt:2012sp,Leslie:2014mua,Ficek:2016qwp,Safronova:2017xyt,Ji:2018tny,Fadeev:2018rfl,Ding:2020mic,Sikivie:2020zpn,
Wu:2021flk,Wei:2022mra,Wang:2022wgs,Wu:2023edi,Wu:2023ypz,Clayburn:2023eys,Huang:2024esg}.  Such studies provide valuable complementary constraints on BSM dynamics, in contrast to large-scale collider experiments..

The interaction potential of the two-body system offers not only a systematical theoretical frame to analyze various laboratory experiments, but also sets a convenient connection between the experimental constraints and the BSM theories.
As the underlying objects of the laboratory measurements at both the atomic and macroscopic scales are basically electrons and nucleons, the interaction potentials arising from the light BSM particle exchanges between two spin-1/2 fermions have been the central focus since the pioneering work in Ref.~\cite{Moody:1984ba}, which calculated the long-range static spin-dependent potentials from the axion exchange.
This formalism was extended in Ref.~\cite{Dobrescu:2006au},  which provides the complete sixteen components of the long-range spin-dependent potentials of the two fermions with the contributions from the spin-0 and spin-1 particles. In the former reference~\cite{Dobrescu:2006au}, a hybrid representation in the momentum and coordinate spaces is employed to describe the spin- and velocity-dependent interaction potentials. Namely the relative velocity $\boldsymbol{v}$ between the two fermions is treated more like a classical vector, instead of an operator as adopted in quantum theory. The hybrid representation of Ref.~\cite{Dobrescu:2006au} is sufficient for the description of the macroscopic experiments, while for the atomic-scale phenomenon it is necessary to treat both coordinate $\boldsymbol{r}$ and momentum (or velocity) $\boldsymbol{p}$ as operators. Accordingly, the hybrid representation formalism is further developed in Ref.~\cite{Fadeev:2018rfl} with the emphasis on the application at the atomic scale. Two main improvements are made for this purpose~\cite{Fadeev:2018rfl}: on one hand the momentum $\boldsymbol{p}$ was promoted as an operator and the potentials are consistently derived in the coordinate space; on the other hand the short-distance effects, characterized by the $\delta(\boldsymbol{r})$ terms, are explicitly worked out and kept in the potentials. Although the latter terms accompanied by $\delta(\boldsymbol{r})$ are not very relevant in the macroscopic phenomenon, they could be crucial for the experiments at the atomic scale.

In this work, we introduce two main improvements over previous studies. First, we derive the relativistic and off-shell corrections to the two-fermion potentials mediated by spin-0 particles, using the Green function rather than the scattering amplitude. While several potential components that vanish in the non-relativistic (NR) limit remain zero in the relativistic case, three terms pick up non-vanishing contributions from the relativistic corrections and seven terms receive contributions due to the off-shell effects. Second, different from the operator basis used in Ref.~\cite{Dobrescu:2006au}, we propose an alternative operator basis that is particularly useful for deriving potentials in coordinate space. The short-distance effects, characterized by $\delta(\boldsymbol{r})$, are explicitly retained in our potentials. As a result, our potentials are applicable to both atomic-scale and macroscopic-scale experimental analyses.

This paper is structured as follows. The relativistic and off-shell corrections to the interaction potentials between two fermions mediated by the exchange of spin-0 particles in the momentum space are derived in detail in Sec.~\ref{sec.momentum}. Both results for the operator bases of Ref.~\cite{Dobrescu:2006au} and ours are analyzed. In Sec.~\ref{sec.coordinate}, we take into account the newly calculated relativistic and off-shell effects to proceed the discussions in the coordinate space. A brief summary and conclusions are given in Sec.~\ref{sec.summary}.

\section{Quasi-potential between two spin-1/2 fermions in momentum space}\label{sec.momentum}

\subsection{Potential in the hybrid representation~\cite{Dobrescu:2006au}}

We consider an elastic scattering process $1+2\to 3 +4$ of two spin-1/2 fermions with four-momentum $p_{i=1,2,3,4}$, where particles 1 and 3 correspond to the same species, and particles 2 and 4 correspond to another same species, i.e., $m_1=m_3=m_a$ and $m_2=m_4=m_b$. In the special case when all the four particles are identical, one would have $m_a=m_b$.
The potential of this process in momentum space, evaluated in the center-of-mass (CM) frame, is expressed as~\cite{Dobrescu:2006au}
\begin{equation}\label{eq.vdm}
\overline{V}(\boldsymbol{q}, \boldsymbol{P})=\mathcal{P}(\boldsymbol{q}^2) \sum_{i=1}^{16} \mathcal{O}^{\RM1}_i(\boldsymbol{q}, \boldsymbol{P}) f_i(\boldsymbol{q}^2, \boldsymbol{P}^2),
\end{equation}
where $\boldsymbol{q}$ and $\boldsymbol{P}$ represent the spatial components of the four-momentum
\begin{equation}
q=p_3-p_1,~~~~P=\frac{1}{2}(p_1+p_3),
\end{equation}
respectively. The function $\mathcal{P}(\boldsymbol{q}^2)$ denotes the denominator of the propagator for the exchanged boson
\begin{equation}
 \mathcal{P}(\boldsymbol{q}^2)= -\frac{1}{\boldsymbol{q}^2+m_0^2}\,,
\end{equation}
with $m_0$ the mass of the exchanged light spin-0 particle.
The operators $\mathcal{O}^{\RM1}_i$ are given in Eqs.(2.2) and (2.3) of Ref.~\cite{Dobrescu:2006au}, and for the sake of completeness we list their explicit expressions
\begin{equation}
\begin{aligned}
\mathcal{O}^{\RM1}_{1} &=1, \quad
\mathcal{O}^{\RM1}_{2} =\boldsymbol{\sigma}^{a} \cdot \boldsymbol{\sigma}^{b}, \quad
\mathcal{O}^{\RM1}_{3} =\frac{1}{m_1^2} (\boldsymbol{\sigma}^{a} \cdot \boldsymbol{q})(\boldsymbol{\sigma}^{b} \cdot \boldsymbol{q}), \quad
\mathcal{O}^{\RM1}_{4,5} =\frac{i}{2m_1^2}(\boldsymbol{\sigma}^{a}\pm \boldsymbol{\sigma}^{b}) \cdot (\boldsymbol{P}\times \boldsymbol{q}),\\
\mathcal{O}^{\RM1}_{6,7} &=\frac{i}{2m_1^2}\left[(\boldsymbol{\sigma}^{a} \cdot \boldsymbol{P})(\boldsymbol{\sigma}^{b} \cdot \boldsymbol{q}) \pm (\boldsymbol{\sigma}^{a} \cdot \boldsymbol{q})(\boldsymbol{\sigma}^{b} \cdot \boldsymbol{P})\right],\qquad
\mathcal{O}^{\RM1}_{8} =\frac{1}{m_1^2} (\boldsymbol{\sigma}^{a} \cdot \boldsymbol{P})(\boldsymbol{\sigma}^{b} \cdot \boldsymbol{P}),
\end{aligned}
\end{equation}
and
\begin{equation}
\begin{aligned}
\mathcal{O}^{\RM1}_{9,10} &=\frac{i}{2m_1}(\boldsymbol{\sigma}^{a} \pm \boldsymbol{\sigma}^{b}) \cdot \boldsymbol{q},\quad
\mathcal{O}^{\RM1}_{11} =\frac{i}{m_1}(\boldsymbol{\sigma}^{a} \times \boldsymbol{\sigma}^{b}) \cdot \boldsymbol{q},\quad
\mathcal{O}^{\RM1}_{12,13} =\frac{1}{2m_1}(\boldsymbol{\sigma}^{a} \pm \boldsymbol{\sigma}^{b}) \cdot \boldsymbol{P},\\
\mathcal{O}^{\RM1}_{14} &=\frac{1}{m_1}(\boldsymbol{\sigma}^{a} \times \boldsymbol{\sigma}^{b}) \cdot \boldsymbol{P},\quad
\mathcal{O}^{\RM1}_{15} =\frac{1}{2m_1^3}\Big\{[\boldsymbol{\sigma}^{a} \cdot (\boldsymbol{P} \times \boldsymbol{q})](\boldsymbol{\sigma}^{b} \cdot \boldsymbol{q})+(\boldsymbol{\sigma}^{a} \cdot \boldsymbol{q})[\boldsymbol{\sigma}^{b} \cdot (\boldsymbol{P} \times \boldsymbol{q})]\Big\},\\
\mathcal{O}^{\RM1}_{16} &=\frac{i}{2m_1^3}\Big\{[\boldsymbol{\sigma}^{a} \cdot (\boldsymbol{P} \times \boldsymbol{q})](\boldsymbol{\sigma}^{b} \cdot \boldsymbol{P})+(\boldsymbol{\sigma}^{a} \cdot \boldsymbol{P})[\boldsymbol{\sigma}^{b} \cdot (\boldsymbol{P} \times \boldsymbol{q})]\Big\}.
\end{aligned}
\end{equation}
The functions $f_i(\boldsymbol{q}^2,\boldsymbol{P}^2)$ in Eq.~\eqref{eq.vdm} are determined from the two-body elastic scattering amplitude.
It should be noted that the labels of the momenta used here differ slightly from those in Ref.~\cite{Dobrescu:2006au}, and there is a sign difference between the  functions $f_i$ presented above and the functions $f_i^{eN}$ in the former reference.

In Ref.~\cite{Dobrescu:2006au}, the potential is derived  from the on-shell condition, leading to the following relations
\begin{equation}
\begin{aligned}
p_i^{0}=&E_i\equiv \sqrt{\boldsymbol{p}_i^2+m_i^2} \,.
\end{aligned}
\end{equation}
Combining these on-shell relations with the four-momentum conservation condition  $p_1+p_2=p_3+p_4$ and the three-momentum relations in the CM frame, i.e., $\boldsymbol{p}_1+\boldsymbol{p}_2=\boldsymbol{p}_3+\boldsymbol{p}_3=\boldsymbol{0}$, one gets the following equalities in the CM frame
\begin{equation}\label{eq.onshellep}
\begin{aligned}
E_{1}=E_{3}, \qquad E_{2}=E_{4}, \qquad  \boldsymbol{q}\cdot\boldsymbol{P}=0\,.
\end{aligned}
\end{equation}
As a result, only two independent variables, $\boldsymbol{P}^2$ and $\boldsymbol{q}^2$, remain in the potential. However, the relations in Eq.~\eqref{eq.onshellep} do not hold any more for off-shell particles. Since the objective electrons or nucleons in the laboratory measurements are not totally free, the on-shell condition can be only considered as an approximation and we will explore the off-shell effects by releasing the on-shell relations in Eq.~\eqref{eq.onshellep} later in this work.

For the exchange of a spin-0 boson $\phi$ between two types of fermions labeled as $a$ and $b$, we consider the following general interaction Lagrangian~\cite{Moody:1984ba,Dobrescu:2006au}
\begin{equation}
\mathcal{L}_\phi=-\phi \bar{\psi}_a(g_{\mathrm{S}}^a+i \gamma_5 g_{\mathrm{P}}^a) \psi_a-\phi \bar{\psi}_b(g_{\mathrm{S}}^{b}+i \gamma_5 g_{\mathrm{P}}^{b}) \psi_b.
\end{equation}

For the electron-nucleon system, i.e., by taking $m_a=m_e$ and $m_b=m_{N}$, the following results are presented in Eq. (6.2) of Ref.~\cite{Dobrescu:2006au}
\begin{equation}
\begin{aligned}
& f_1^{e N}(0,0)=-g_{\mathrm{S}}^e g_{\mathrm{S}}^{N}, \\
& f_3^{e N}(0,0)=-\frac{m_e}{4 m_{N}} g_{\mathrm{P}}^e g_{\mathrm{P}}^{N}, \\
& f_{4,5}^{e N}(0,0)=-\frac{1}{4}(1 \pm \frac{m_e^2}{m_{N}^2}) g_{\mathrm{S}}^e g_{\mathrm{S}}^{N}, \\
& f_{9,10}^{e N}(0,0)=\frac{1}{2}(g_{\mathrm{P}}^e g_{\mathrm{S}}^{N} \mp g_{\mathrm{S}}^e g_{\mathrm{P}}^{N} \frac{m_e}{m_{N}}), \\
& f_{15}^{e N}(0,0)=\frac{m_e}{4 m_{N}}(g_{\mathrm{S}}^e g_{\mathrm{P}}^{N}-g_{\mathrm{P}}^e g_{\mathrm{S}}^{N} \frac{m_e}{m_{N}}).
\end{aligned}
\end{equation}
These formulas are obtained by taking the NR limit and the on-shell condition. Next, we generalize these results by including the relativistic and off-shell corrections.

\subsection{Derivation of potential from Bethe-Salpeter equation}

In principle, the quasi-potential in the $\mathrm{Schr\ddot{o}dinger}$-like equation is determined by the interaction kernel in the four-dimensional Bethe-Salpeter (BS) equation. Within this framework, total four-momentum conservation holds, while the momenta of the incoming and outgoing particles may be defined as off-shell, indicating that $p_i^{0}\neq E_{i}$. This can lead to $\boldsymbol{q}\cdot \boldsymbol{P}\neq 0$, comparing with the on-shell case in Eq.~\eqref{eq.onshellep}. Therefore one would expect that the interaction potentials should also contain additional terms with $\boldsymbol{q}\cdot \boldsymbol{P}$, which are however absent in Ref.~\cite{Dobrescu:2006au} due to the assumption of on-shell condition.

In the CM frame, the exact quasi-potential for an elastic two-fermion scattering process $1+2\to 3 +4$, is derived as \cite{Caswell:1978mt}
\begin{equation}
\begin{aligned}
&\chi^{\dag}_{\bar{i}}(\lambda_3)\chi^{\dag}_{\bar{j}}(\lambda_4)\overline{V}_{\bar{i}i,\bar{j}j}(\boldsymbol{q}, \boldsymbol{P})\chi_{i}(\lambda_1)\chi_{j}(\lambda_2)&\equiv i\widetilde{K}_{\lambda_3\lambda_1, \lambda_4\lambda_2}(\boldsymbol{q}, \boldsymbol{P}, \sqrt{s}),
\end{aligned}
\end{equation}
where $\chi$ represents the Pauli spinor, $\lambda_i$ denote the helicities and $s\equiv(p_1+p_2)^2=(E_1+E_2)^2$. The expression for $\widetilde{K}_{\lambda_3\lambda_1, \lambda_4\lambda_2}(\boldsymbol{q}, \boldsymbol{P}, \sqrt{s})$ is given by
\begin{equation}
\begin{aligned}
&\widetilde{K}_{\lambda_3\lambda_1, \lambda_4\lambda_2}(\boldsymbol{q}, \boldsymbol{P}, \sqrt{s})\equiv \nonumber \\
& \qquad \frac{\bar{u}_{\bar{\alpha}}(\boldsymbol{p}_{3},m_3,\lambda_3) \bar{u}_{\bar{\beta}}(\boldsymbol{p}_{4},m_4,\lambda_4)}{\sqrt{4 E_{3} E_{4}}}
\bar{K}_{\bar{\alpha}\alpha,\bar{\beta}\beta}(\boldsymbol{q}, \boldsymbol{P}, \sqrt{s}) \frac{u_{\alpha}(\boldsymbol{p}_{1},m_1,\lambda_1) u_{\beta}(\boldsymbol{p}_{2}, m_2,\lambda_2)}{\sqrt{4 E_{1}E_{2}}}\,,
\end{aligned}
\end{equation}
where $u(\boldsymbol{p}_i, m_i,\lambda_i)$ is the Dirac spinor normalized as
\begin{equation}
\begin{aligned}
\bar{u}(\boldsymbol{p}_i, m_i,\lambda_i)u(\boldsymbol{p}_i, m_i,\lambda_i)=2m_i.
\end{aligned}
\end{equation}
Additionally, $\bar{K}(\boldsymbol{q}, \boldsymbol{P}, \sqrt{s})$ is defined as
\begin{equation}
\begin{aligned}
\bar{K}(\boldsymbol{q}, \boldsymbol{P}, \sqrt{s})&\equiv \bar{K}(q, P,  p_1+p_2)\Big|_{p_{1}^{0}=p_{3}^{0}=\tau \sqrt{s}},\\
\bar{K} &\equiv [I-K_{\textrm{BS}}(G_0-\bar{G}_0)]^{-1} K_{\textrm{BS}},
\end{aligned}
\end{equation}
where $\tau\equiv \frac{m_1}{m_1+m_2}$, $I$ is the identity matrix, $K_{\mathrm{BS}}$ is the interaction kernel in the conventional four-dimensional BS equation, and
\begin{equation}
\begin{aligned}
G_0(p_1, p_2)&\equiv S^{(a)}_{full}(p_1,m_a)S^{(b)}_{full}(p_2,m_{b}),\\
\bar{G}_0(p_1, p_2,q)&\equiv 2 \pi i \delta(q^0)\Big[\frac{\Lambda_{+}^{(a)}(\boldsymbol{p}_1) \Lambda_{+}^{(b)}(-\boldsymbol{p}_1)}{\sqrt{s}-E_1-E_{3}}-\frac{\Lambda_{-}^{(a)}(\boldsymbol{p}_1) \Lambda_{-}^{(b)}(-\boldsymbol{p}_2)}{\sqrt{s}+E_1+E_3}\Big]\\
&\approx 2 \pi i \delta(q^0)\frac{\Lambda_{+}^{(a)}(\boldsymbol{p}_1) \Lambda_{+}^{(b)}(-\boldsymbol{p}_1)}{\sqrt{s}-E_1-E_3}.
\end{aligned}
\end{equation}
Here, $S^{(a,b)}_{full}$ are the full propagators of the particles 1 and 2 with masses $m_a$ and $m_b$~\cite{Jentschura:2022xuc}, respectively, and $\Lambda_{\pm}^{(a,b)}(\boldsymbol{k})$ are
\begin{equation}
\begin{aligned}
\Lambda_{\pm}^{(a,b)}(\boldsymbol{k})&\equiv\left[E_{a,b}(\boldsymbol{k}) \gamma^0 \mp(\boldsymbol{k} \cdot \vec{\gamma}-m_{a,b})\right] / 2 E_{a,b}(\boldsymbol{k}),\\
E_{a,b}(\boldsymbol{k}) &\equiv \sqrt{\boldsymbol{k}^2+m_{a,b}^2} .
\end{aligned}
\end{equation}

In the aforementioned CM frame, since both $\sqrt{s}$ and $p_{1,3}^{0}$ are fixed, three independent variables remain, which will be taken as $\boldsymbol{q}^2, \boldsymbol{P}^2$ and $\boldsymbol{q}\cdot\boldsymbol{P}$. The  general potential in the momentum space can be then expressed as
\begin{equation}
\overline{V}(\boldsymbol{q}, \boldsymbol{P})=\mathcal{P}(\boldsymbol{q}^2) \sum_{i=1}^{16} \mathcal{O}^{\RM1}_i(\boldsymbol{q}, \boldsymbol{P}) f_i(\boldsymbol{q}^2, \boldsymbol{P}^2,\boldsymbol{q}\cdot\boldsymbol{P}).
\end{equation}
At leading order (LO) in the couplings $g_{\mathrm{S,P}}^{a,b}$, one has
\begin{equation}
\begin{aligned}
\bar{K}_{\bar{\alpha}\alpha,\bar{\beta}\beta}(q, P,  p_1+p_2)\Big|_{p_{1}^{0}=p_{3}^{0}=\tau \sqrt{s}}
&\approx K_{\textrm{BS}}(q, P, p_1+p_2)\Big|_{p_{1}^{0}=p_{3}^{0}=\tau \sqrt{s}} \\
=& i (g_{\mathrm{S}}^{a}+ig_{\mathrm{P}}^{a}\gamma_5)_{\bar{\alpha}\alpha}(g_{\mathrm{S}}^{b}+ig_{\mathrm{P}}^{b}\gamma_5)_{\bar{\beta}\beta}\mathcal{P}(\boldsymbol{q}^2)\,.
\end{aligned}
\end{equation}
By utilizing these relations, the coefficients of the quasi-potential at LO in the couplings, accounting for the relativistic effects and the off-shell terms, can be written as follows:
\begin{equation}
\begin{aligned}
f_1^{\textrm{full}}=&-\frac{1}{16}g_{\mathrm{S}}^{a}g_{\mathrm{S}}^{b}FY_1Y_2,\\
f_2^{\textrm{full}}=&g_{\mathrm{S}}^{a} g_{\mathrm{S}}^{b}F [\boldsymbol{P}^2 \boldsymbol{q}^2-(\boldsymbol{q}\cdot \boldsymbol{P})^2],\\
f_3^{\textrm{full}}=&-g_{\mathrm{S}}^{a} g_{\mathrm{S}}^{b}F m_{a}^2 \boldsymbol{P}^2-\frac{1}{4} g_{\mathrm{P}}^{a} g_{\mathrm{P}}^{b}Fm_{a}^2(X_{1}+\mathrm{X_{3}})(X_{2}+X_{4}),\\
f_4^{\textrm{full}}=&-\frac{1}{4} g_{\mathrm{S}}^{a} g_{\mathrm{S}}^{b}Fm_{a}^2(Y_1+Y_2),\\
f_5^{\textrm{full}}=&g_{\mathrm{S}}^{a} g_{\mathrm{S}}^{b}Fm_{a}^2(X_{1}X_{3}-X_{2} X_{4}),\\
f_6^{\textrm{full}}=&-2 i g_{\mathrm{S}}^{a} g_{\mathrm{S}}^{b}Fm_{a}^2(\boldsymbol{q}\cdot \boldsymbol{P})+ i g_{\mathrm{P}}^{a} g_{\mathrm{P}}^{b}m_{a}^2F(E_{1}X_{2}+E_{2}m_{a}-E_{3}X_{4}-E_{4}m_{a}),\\
f_7^{\textrm{full}}=&i g_{\mathrm{P}}^{a} g_{\mathrm{P}}^{b}Fm_{a}^2(E_{1}X_{4}-E_{2} X_{3}-E_{3} m_{b}+E_{4}m_{a}),\\
f_8^{\textrm{full}}=&-g_{\mathrm{S}}^{a} g_{\mathrm{S}}^{b}Fm_{a}^2 \boldsymbol{q}^2-g_{\mathrm{P}}^{a}g_{\mathrm{P}}^{b}Fm_{a}^2(E_{1}-E_{3})(E_{2}-E_{4}),\\
\end{aligned}
\end{equation}
and
\begin{equation}
\begin{aligned}
f_9^{\textrm{full}}=&-\frac{1}{8} g_{\mathrm{S}}^{a}g_{\mathrm{P}}^{b}Fm_{a}(X_{2}+X_{4})Y_1+\frac{1}{8} g_{\mathrm{P}}^{a}g_{\mathrm{S}}^{b}Fm_{a}(X_{1}+X_{3})Y_2,\\
f_{10}^{\textrm{full}}=&\frac{1}{8} g_{\mathrm{S}}^{a}g_{\mathrm{P}}^{b}Fm_{a}(X_{2}+X_{4})Y_1+ \frac{1}{8} g_{\mathrm{P}}^{a}g_{\mathrm{S}}^{b}Fm_{a}(X_{1}+X_{3})Y_2,\\
f_{11}^{\textrm{full}}=&-\frac{1}{8}  g_{\mathrm{S}}^{a}g_{\mathrm{P}}^{b}Fm_{a}(E_{2}-E_{4})Y_1+\frac{1}{8}  g_{\mathrm{P}}^{a} g_{\mathrm{S}}^{b}Fm_a(E_{1}-E_{3})Y_2,\\
f_{12}^{\textrm{full}}=&\frac{1}{4} i g_{\mathrm{S}}^{a}g_{\mathrm{P}}^{b}Fm_{a}(E_{2}-E_{4})Y_1+\frac{1}{4} i g_{\mathrm{P}}^{a} g_{\mathrm{S}}^{b}Fm_a(E_{1}-E_{3})Y_2,\\
f_{13}^{\textrm{full}}=&\frac{1}{2} g_{\mathrm{S}}^{a}g_{\mathrm{P}}^{b}Fm_aZ_1+\frac{1}{2} g_{\mathrm{P}}^{a} g_{\mathrm{S}}^{b}Fm_aZ_2,\\
f_{14}^{\textrm{full}}=&-\frac{1}{4} g_{\mathrm{S}}^{a}g_{\mathrm{P}}^{b}Fm_aZ_3-\frac{1}{4} g_{\mathrm{P}}^{a} g_{\mathrm{S}}^{b}Fm_aZ_4,\\
f_{15}^{\textrm{full}}=&\frac{1}{2} g_{\mathrm{S}}^{a}g_{\mathrm{P}}^{b} Fm_{a}^3(X_{2}+X_{4})-\frac{1}{2} g_{\mathrm{P}}^{a} g_{\mathrm{S}}^{b}Fm_{a}^3(X_{1}+X_{3}),\\
f_{16}^{\textrm{full}}=&-ig_{\mathrm{S}}^{a}g_{\mathrm{P}}^{b}Fm_a^3(E_{2}-E_{4})+ig_{\mathrm{P}}^{a}g_{\mathrm{S}}^{b}Fm_a^3(E_{1}-E_{3}),
\end{aligned}
\end{equation}
with
\begin{equation}
\begin{aligned}
X_i&\equiv E_i+m_i,\\
F&\equiv \frac{1}{4\sqrt{X_1X_2X_3X_4E_1E_2E_3E_4}},\\
Y_1&\equiv -4 \boldsymbol{P}^2+\boldsymbol{q}^2+4X_{1}\mathrm{X_{3}},\\
Y_2&\equiv-4 \boldsymbol{P}^2+\boldsymbol{q}^2+4 X_{2} X_{4},\\
Z_1&\equiv2\boldsymbol{P}^2(E_{2}-E_{4})+\boldsymbol{q}\cdot\boldsymbol{P}(X_2+X_4),\\
Z_2&\equiv2\boldsymbol{P}^2(E_{1}-E_{3})+\boldsymbol{q}\cdot\boldsymbol{P}(X_{1}+X_{3}),\\
Z_3&\equiv2(E_{2} -E_{4})\boldsymbol{q}\cdot \boldsymbol{P}+\boldsymbol{q}^2(X_{2}+X_{4}),\\
Z_4&\equiv2(E_{1}-E_{3})\boldsymbol{q}\cdot \boldsymbol{P}+\boldsymbol{q}^2(X_{1}+X_{3}).
\end{aligned}
\end{equation}

To make an easy comparison with the results in the non-relativistic (NR) limit presented in Ref.~\cite{Dobrescu:2006au}, we expand the above coefficients in terms of the small momenta $\boldsymbol{P}$ and $\boldsymbol{q}$, while not imposing the on-shell conditions. Then the contributions can be separated into the on-shell (containing the constant, $\boldsymbol{q}^2$ and $\boldsymbol{P}^2$ factors) and off-shell (containing the $\boldsymbol{q} \cdot \boldsymbol{P}$ factor) parts
\begin{equation}\label{eq.vinrm}
 \overline{V}(\boldsymbol{q}, \boldsymbol{P})\rightarrow \overline{V}^{\RM1,\text{NR}}=\mathcal{P}(\boldsymbol{q}^2) \sum_{i=1}^{16} \mathcal{O}^{\RM1}_i(\boldsymbol{q}, \boldsymbol{P})[ f_{i,\textrm{on}}^{\textrm{NR}}(\boldsymbol{q}^2, \boldsymbol{P}^2)+ f_{i,\textrm{off}}^{\textrm{NR}}(\boldsymbol{q}\cdot\boldsymbol{P})],
\end{equation}
where the nonzero on-shell terms read
\begin{equation}\label{eq.vinrmf1}
\begin{aligned}
f_{1,\textrm{on}}^{\textrm{NR}}=&-g_{\mathrm{S}}^{a} g_{\mathrm{S}}^{b} \left(1-\frac{\boldsymbol{P}^2}{2 m_{a}^2}-\frac{\boldsymbol{P}^2}{2 m_{b}^2}\right),\\
f_{2,\textrm{on}}^{\textrm{NR}}=&g_{\mathrm{S}}^{a} g_{\mathrm{S}}^{b} \frac{\boldsymbol{P}^2 \boldsymbol{q}^2}{16 m_{a}^2 m_{b}^2},\\
f_{3,\textrm{on}}^{\textrm{NR}}=&-g_{\mathrm{S}}^{a} g_{\mathrm{S}}^{b} \frac{\boldsymbol{P}^2}{16 m_{b}^2}-g_{\mathrm{P}}^{a} g_{\mathrm{P}}^{b} \left(\frac{m_{a}}{4 m_{b}}-\frac{m_{a} \boldsymbol{P}^2 }{8 m_{b}^3}-\frac{m_{a} \boldsymbol{q}^2 }{32 m_{b}^3}-\frac{\boldsymbol{P}^2}{8 m_{a}m_{b}}-\frac{\boldsymbol{q}^2}{32 m_{a}m_{b}}\right),\\
f_{4,\textrm{on}}^{\textrm{NR}}=&-g_{\mathrm{S}}^{a} g_{\mathrm{S}}^{b} \left(\frac{1}{4}+\frac{m_{a}^2}{4 m_{b}^2}-\frac{3 m_{a}^2 \boldsymbol{P}^2 }{16 m_{b}^4}-\frac{3 m_{a}^2 \boldsymbol{q}^2 }{64 m_{b}^4}-\frac{\boldsymbol{P}^2}{4 m_{b}^2}-\frac{3 \boldsymbol{P}^2}{16 m_{a}^2}-\frac{3 \boldsymbol{q}^2}{64 m_{a}^2}\right),\\
f_{5,\textrm{on}}^{\textrm{NR}}=&-g_{\mathrm{S}}^{a} g_{\mathrm{S}}^{b} \left(\frac{1}{4}-\frac{m_{a}^2}{4 m_{b}^2}+\frac{3 m_{a}^2 \boldsymbol{P}^2 }{16 m_{b}^4}+\frac{3 m_{a}^2 \boldsymbol{q}^2 }{64 m_{b}^4}-\frac{3 \boldsymbol{P}^2}{16 m_{a}^2}-\frac{3 \boldsymbol{q}^2}{64 m_{a}^2}\right),\\
f_{8,\textrm{on}}^{\textrm{NR}}=&-g_{\mathrm{S}}^{a} g_{\mathrm{S}}^{b} \frac{\boldsymbol{q}^2}{16 m_{b}^2},\\
f_{9,\textrm{on}}^{\textrm{NR}}=&-g_{\mathrm{S}}^{a}g_{\mathrm{P}}^{b}  \left(\frac{m_{a}}{2 m_{b}}-\frac{m_{a} \boldsymbol{P}^2 }{4 m_{b}^3}-\frac{m_{a} \boldsymbol{q}^2 }{16 m_{b}^3}-\frac{\boldsymbol{P}^2}{4 m_{a}m_{b}}\right)+g_{\mathrm{P}}^{a} g_{\mathrm{S}}^{b} \left(\frac{1}{2}-\frac{\boldsymbol{P}^2}{4 m_{a}^2}-\frac{\boldsymbol{P}^2}{4 m_{b}^2}-\frac{\boldsymbol{q}^2}{16 m_{a}^2}\right),\\
f_{10,\textrm{on}}^{\textrm{NR}}=&g_{\mathrm{S}}^{a}g_{\mathrm{P}}^{b}  \left(\frac{m_{a}}{2 m_{b}}-\frac{m_{a} \boldsymbol{P}^2 }{4 m_{b}^3}-\frac{m_{a} \boldsymbol{q}^2 }{16 m_{b}^3}-\frac{\boldsymbol{P}^2}{4 m_{a}m_{b}}\right)+g_{\mathrm{P}}^{a} g_{\mathrm{S}}^{b} \left(\frac{1}{2}-\frac{\boldsymbol{P}^2}{4 m_{a}^2}-\frac{\boldsymbol{P}^2}{4 m_{b}^2}-\frac{\boldsymbol{q}^2}{16 m_{a}^2}\right),\\
f_{14,\textrm{on}}^{\textrm{NR}}=&-g_{\mathrm{S}}^{a} g_{\mathrm{P}}^{b}\frac{ \boldsymbol{q}^2}{16 m_{a} m_{b}}-g_{\mathrm{P}}^{a} g_{\mathrm{S}}^{b} \frac{\boldsymbol{q}^2}{16 m_{b}^2},\\
f_{15,\textrm{on}}^{\textrm{NR}}=&g_{\mathrm{S}}^{a}g_{\mathrm{P}}^{b} \frac{ m_{a}}{8 m_{b}}-g_{\mathrm{P}}^{a} g_{\mathrm{S}}^{b}\frac{ m_{a}^2}{8 m_{b}^2},
\end{aligned}
\end{equation}
and the nonzero off-shell terms are
\begin{equation}\label{eq.vinrmf2}
\begin{aligned}
 f_{1,\textrm{off}}^{\textrm{NR}}=&g_{\mathrm{S}}^{a} g_{\mathrm{S}}^{b} \left(-\frac{5 (\boldsymbol{q} \cdot \boldsymbol{P})^2}{32 m_{a}^4}-\frac{5 (\boldsymbol{q} \cdot \boldsymbol{P})^2}{32 m_{b}^4}\right),\\
f_{2,\textrm{off}}^{\textrm{NR}}=&-g_{\mathrm{S}}^{a} g_{\mathrm{S}}^{b}\frac{ (\boldsymbol{q} \cdot \boldsymbol{P})^2}{16 m_{a}^2 m_{b}^2},\\
 f_{6,\textrm{off}}^{\textrm{NR}}=&-g_{\mathrm{S}}^{a} g_{\mathrm{S}}^{b}\frac{i \boldsymbol{q} \cdot \boldsymbol{P}}{8 m_{b}^2}+g_{\mathrm{P}}^{a} g_{\mathrm{P}}^{b} \left(-\frac{i \boldsymbol{q} \cdot \boldsymbol{P}}{8 m_{a} m_{b}}-\frac{i m_{a} \boldsymbol{q} \cdot \boldsymbol{P}}{8 m_{b}^3}\right),\\
f_{7,\textrm{off}}^{\textrm{NR}}=&g_{\mathrm{P}}^{a} g_{\mathrm{P}}^{b} \left(\frac{i m_{a} \boldsymbol{q} \cdot \boldsymbol{P}}{8 m_{b}^3}-\frac{i \boldsymbol{q} \cdot \boldsymbol{P}}{8 m_{a} m_{b}}\right),\\
f_{11,\textrm{off}}^{\textrm{NR}}=&g_{\mathrm{S}}^{a}g_{\mathrm{P}}^{b}\frac{  m_{a}\boldsymbol{q} \cdot \boldsymbol{P}}{8 m_{b}^3}-g_{\mathrm{P}}^{a} g_{\mathrm{S}}^{b}\frac{ \boldsymbol{q} \cdot \boldsymbol{P}}{8 m_{a}^2},\\
f_{12,\textrm{off}}^{\textrm{NR}}=&-g_{\mathrm{S}}^{a}g_{\mathrm{P}}^{b}\frac{i   m_{a} \boldsymbol{q} \cdot \boldsymbol{P}}{4 m_{b}^3}-g_{\mathrm{P}}^{a} g_{\mathrm{S}}^{b}\frac{i\boldsymbol{q} \cdot \boldsymbol{P}}{4 m_{a}^2},\\
f_{13,\textrm{off}}^{\textrm{NR}}=&g_{\mathrm{S}}^{a}g_{\mathrm{P}}^{b}\frac{ \boldsymbol{q} \cdot \boldsymbol{P}}{8 m_{a} m_{b}}+g_{\mathrm{P}}^{a} g_{\mathrm{S}}^{b}\frac{\boldsymbol{q} \cdot \boldsymbol{P}}{8 m_{b}^2},
\end{aligned}
\end{equation}
with all other terms equal to zero. In the above expressions,  most of the terms are expanded up to the next-leading order (NLO) with respect to the small momenta, except for $f_{2,\textrm{on}}^{\textrm{NR}}$, $f_{1,\textrm{off}}^{\textrm{NR}}$ and $f_{2,\textrm{off}}^{\textrm{NR}}$, which are expanded up to the next-to-next-leading-order (NNLO), because these latter three terms vanish at LO and NLO.

By neglecting the terms $\boldsymbol{P}^2$, $\boldsymbol{q}^2$ and $\boldsymbol{q}\cdot\boldsymbol{P}$, the above results in Eqs.~\eqref{eq.vinrmf1} and \eqref{eq.vinrmf2} exactly reduce to the formulas presented in Ref.~\cite{Dobrescu:2006au}, except for a global factor $\frac{1}{2}$ difference in $f_{15}$. The new contributions arising from the terms with  $\boldsymbol{P}^2$ and $\boldsymbol{q}^2$ account for the NLO or NNLO relativistic contributions, in the meanwhile off-shell effects are also taken into account by the $\boldsymbol{q}\cdot\boldsymbol{P}$ terms.

\subsection{Potential in another form suitable for both momentum and coordinate spaces}\label{sec.newopm}

As pointed out in Ref.~\cite{Fadeev:2018rfl}, it is not straightforward to match the potential form of Ref.~\cite{Dobrescu:2006au} into the coordinate space when promoting the momentum as an operator. It is noted that for atomic systems there is a simple recipe to overcome this issue by choosing proper kinematical variables. E.g., typically one could take $\boldsymbol{q}$ and $\boldsymbol{p}_1$ (the three-momentum of particle 1) \cite{Berestetskii:1982qgu}, instead of $\boldsymbol{q}$ and $\boldsymbol{P}$, to conveniently relate the potentials in the momentum and coordinate spaces. Therefore we propose a different form to express the potential in momentum space as
\begin{equation}\label{eq.vii}
\overline{V}(\boldsymbol{q}, \boldsymbol{P})=\mathcal{P}(\boldsymbol{q}^2) \sum_{i=1}^{16} \mathcal{O}^{\RM2}_i(\boldsymbol{q}, \boldsymbol{p}_1) g_i(\boldsymbol{q}, \boldsymbol{p}_1),
\end{equation}
where the three-momenta $\boldsymbol{p}_1$ and $\boldsymbol{q}$ are utilized.
Here we introduce the superscript $\RM2$ for the operators $\mathcal{O}^{\RM2}_{i}$, in order to distinguish the operator basis of Ref.~\cite{Dobrescu:2006au}, labeled as $\mathcal{O}^{\rm I}_{i}$ in this work.  The operator basis $\mathcal{O}^{\RM2}_{i}$ can be separated into three types, spin-independent, one-spine-dependent, and two-spine-dependent. The spin-independent and one-spin-dependent operators are chosen as
\begin{equation}\label{eq.oii1}
\begin{aligned}
\mathcal{O}^{\RM2}_{1} &\equiv 1,\\
\mathcal{O}^{\RM2}_{2} &\equiv \frac{i}{\mu^2} \boldsymbol{\sigma}^{a}\cdot \boldsymbol{Q},
&\mathcal{O}^{\RM2}_{3} &\equiv \frac{i}{\mu^2}\boldsymbol{\sigma}^{b}\cdot \boldsymbol{Q}, \\
\mathcal{O}^{\RM2}_{4}&\equiv \frac{i}{\mu}\boldsymbol{\sigma}^{a}\cdot \boldsymbol{q},
&\mathcal{O}^{\RM2}_{5} &\equiv \frac{i}{\mu}\boldsymbol{\sigma}^{b}\cdot \boldsymbol{q},\\
\mathcal{O}^{\RM2}_{6} &\equiv \frac{i}{\mu} \boldsymbol{\sigma}^{a}\cdot \boldsymbol{p}_1,
&\mathcal{O}^{\RM2}_{7} &\equiv \frac{i}{\mu}\boldsymbol{\sigma}^{b}\cdot \boldsymbol{p}_1,\\
\end{aligned}
\end{equation}
and the two-spin-dependent operators are chosen as
\begin{equation}\label{eq.oii2}
\begin{aligned}
\mathcal{O}^{\RM2}_{8}&\equiv \boldsymbol{\sigma}^{a}\cdot \boldsymbol{\sigma}^{b},\\
\mathcal{O}^{\RM2}_{9} &\equiv \frac{1}{\mu^2}\boldsymbol{\sigma}^{a}\cdot \boldsymbol{q} \boldsymbol{\sigma}^{b}\cdot \boldsymbol{q},
&\mathcal{O}^{\RM2}_{10}&\equiv \frac{1}{\mu^2} \boldsymbol{\sigma}^{a}\cdot \boldsymbol{p}_1 \boldsymbol{\sigma}^{b}\cdot \boldsymbol{q},\\
\mathcal{O}^{\RM2}_{11}&\equiv \frac{1}{\mu^2}\boldsymbol{\sigma}^{a}\cdot \boldsymbol{q} \boldsymbol{\sigma}^{b}\cdot \boldsymbol{p}_1,
&\mathcal{O}^{\RM2}_{12}&\equiv \frac{1}{\mu^2}\boldsymbol{\sigma}^{a}\cdot \boldsymbol{p}_1 \boldsymbol{\sigma}^{b}\cdot \boldsymbol{p}_1,\\
\mathcal{O}^{\RM2}_{13}&\equiv \frac{1}{\mu^3}\boldsymbol{\sigma}^{a}\cdot \boldsymbol{Q} \boldsymbol{\sigma}^{b}\cdot \boldsymbol{q},
&\mathcal{O}^{\RM2}_{14}&\equiv \frac{1}{\mu^3}\boldsymbol{\sigma}^{a}\cdot \boldsymbol{Q} \boldsymbol{\sigma}^{b}\cdot \boldsymbol{p}_1, \\
\mathcal{O}^{\RM2}_{15}&\equiv\frac{1}{\mu}(\boldsymbol{\sigma}^{a}\times \boldsymbol{\sigma}^{b}) \cdot \boldsymbol{q},
&\mathcal{O}^{\RM2}_{16}&\equiv\frac{1}{\mu}(\boldsymbol{\sigma}^{a}\times \boldsymbol{\sigma}^{b}) \cdot \boldsymbol{p}_1,
\end{aligned}
\end{equation}
with $\mu\equiv \frac{m_am_{b}}{m_a+m_{b}}$ and $\boldsymbol{Q}\equiv\boldsymbol{q}\times\boldsymbol{p}_1$. In principle, one could also construct additional forms for the operators, such as
\begin{equation}
\begin{aligned}
\mathcal{O}^{\RM2}_{17}&\equiv \frac{1}{\mu^3}\boldsymbol{\sigma}^{a}\cdot \boldsymbol{q}\boldsymbol{\sigma}^{b}\cdot \boldsymbol{Q}, \\
\mathcal{O}^{\RM2}_{18}&\equiv \frac{1}{\mu^3}\boldsymbol{\sigma}^{a}\cdot \boldsymbol{p}_1\boldsymbol{\sigma}^{b}\cdot \boldsymbol{Q},
\end{aligned}
\end{equation}
which are however not independent from the above sixteen terms due to the following relations
\begin{equation}
\begin{aligned}
\mathcal{O}^{\RM2}_{17}=&\mathcal{O}^{\RM2}_{13}+\frac{1}{\mu^2}\boldsymbol{q}\cdot\boldsymbol{p}_1 \mathcal{O}^{\RM2}_{15} -\frac{1}{\mu^2}\boldsymbol{q}^2\mathcal{O}^{\RM2}_{16}, \\
\mathcal{O}^{\RM2}_{18}=&\mathcal{O}^{\RM2}_{14}+\frac{1}{\mu^2}\boldsymbol{p}_1^2\mathcal{O}^{\RM2}_{15} -\frac{1}{\mu^2}\boldsymbol{q}\cdot\boldsymbol{p}_1 \mathcal{O}^{\RM2}_{16}.
\end{aligned}
\end{equation}

Similar with calculation in the last subsection, the coefficients $g_i$ of the quasi-potential~\eqref{eq.vii} at the LO of couplings and with the full relativistic and off-shell effects, can be expressed as
\begin{equation}\label{eq.gir1}
\begin{aligned}
g_1^{\textrm{full}}=&-g_{\mathrm{S}}^a g_{\mathrm{S}}^{b} F(X_{1} X_{3}-\boldsymbol{p}_1^2-\boldsymbol{q} \cdot \boldsymbol{p}_1) (X_{2} X_{4}-\boldsymbol{p}_1^2-\boldsymbol{q} \cdot \boldsymbol{p}_1),\\
g_2^{\textrm{full}}=&g_{\mathrm{S}}^a g_{\mathrm{S}}^{b} F\mu^2 (X_{2} X_{4}-\boldsymbol{p}_1^2-\boldsymbol{q} \cdot \boldsymbol{p}_1),\\
g_3^{\textrm{full}}=&g_{\mathrm{S}}^a g_{\mathrm{S}}^{b} F\mu^2 (X_{1} X_{3}-\boldsymbol{p}_1^2-\boldsymbol{q} \cdot \boldsymbol{p}_1),\\
g_4^{\textrm{full}}=& g_{\mathrm{P}}^{a} g_{\mathrm{S}}^{b}F \mu X_{1} (X_{2} X_{4}-\boldsymbol{p}_1^2-\boldsymbol{q} \cdot \boldsymbol{p}_1),\\
g_5^{\textrm{full}}=&-g_{\mathrm{S}}^a g_{\mathrm{P}}^{b}F \mu  X_{2} (X_{1} X_{3}-\boldsymbol{p}_1^2-\boldsymbol{q} \cdot \boldsymbol{p}_1),\\
g_6^{\textrm{full}}=&  g_{\mathrm{P}}^{a} g_{\mathrm{S}}^{b}F \mu (E_{1}-E_{3})(X_{2} X_{4}-\boldsymbol{p}_1^2-\boldsymbol{q} \cdot \boldsymbol{p}_1),\\
g_7^{\textrm{full}}=&-  g_{\mathrm{S}}^a  g_{\mathrm{P}}^{b}F \mu(E_{2}-E_{4}) (X_{1} X_{3}-\boldsymbol{p}_1^2-\boldsymbol{q} \cdot \boldsymbol{p}_1),\\
g_8^{\textrm{full}}=&g_{\mathrm{S}}^a g_{\mathrm{S}}^{b}F [\boldsymbol{p}_1^2 \boldsymbol{q}^2-(\boldsymbol{q} \cdot \boldsymbol{p}_1)^2],
\end{aligned}
\end{equation}
and
\begin{equation}\label{eq.gir2}
\begin{aligned}
g_9^{\textrm{full}}=& -g_{\mathrm{P}}^{a} g_{\mathrm{P}}^{b}F\mu^2 X_{1} X_{2}-g_{\mathrm{S}}^a g_{\mathrm{S}}^{b} F\mu^2 \boldsymbol{p}_1^2,\\
g_{10}^{\textrm{full}}=& g_{\mathrm{S}}^a g_{\mathrm{S}}^{b}F \mu^2 \boldsymbol{q} \cdot \boldsymbol{p}_1-g_{\mathrm{P}}^{a} g_{\mathrm{P}}^{b}F\mu^2 X_{2}(E_{1}-E_{3}),\\
g_{11}^{\textrm{full}}=& g_{\mathrm{S}}^a g_{\mathrm{S}}^{b}F \mu^2 \boldsymbol{q} \cdot \boldsymbol{p}_1-g_{\mathrm{P}}^{a} g_{\mathrm{P}}^{b}F\mu^2 X_{1}(E_{2}-E_{4}),\\
g_{12}^{\textrm{full}}=& -g_{\mathrm{S}}^a g_{\mathrm{S}}^{b}F \mu^2 \boldsymbol{q}^2-g_{\mathrm{P}}^{a} g_{\mathrm{P}}^{b}F\mu^2(E_{1}-E_{3}) (E_{2}-E_{4}),\\
g_{13}^{\textrm{full}}=&  -g_{\mathrm{S}}^a g_{\mathrm{P}}^{b}F\mu^3 X_{2}+g_{\mathrm{P}}^a g_{\mathrm{S}}^{b}F\mu^3 X_{1},\\
g_{14}^{\textrm{full}}=&-g_{\mathrm{S}}^{a} g_{\mathrm{P}}^{b}F\mu^3 (E_{2}-E_{4})+g_{\mathrm{P}}^{a} g_{\mathrm{S}}^{b}F \mu^3 (E_{1}-E_{3}),\\
g_{15}^{\textrm{full}}=&  g_{\mathrm{P}}^a g_{\mathrm{S}}^{b}F\mu [(E_{1}-E_{3}) \boldsymbol{p}_1^2+X_{1} \boldsymbol{q} \cdot \boldsymbol{p}_1],\\
g_{16}^{\textrm{full}}=&- g_{\mathrm{P}}^{a} g_{\mathrm{S}}^{b}F\mu [(E_{1}-E_{3}) \boldsymbol{q} \cdot \boldsymbol{p}_1+X_{1} \boldsymbol{q}^2].
\end{aligned}
\end{equation}

After expanding these coefficients in terms of the small three-momenta $\boldsymbol{p}_1$ and $\boldsymbol{q}$, we obtain
\begin{equation}
V(\boldsymbol{q}, \boldsymbol{p}_1)\rightarrow V^{\RM2,\text{NR}}=\mathcal{P}(\boldsymbol{q}^2) \sum_{i=1}^{16} \mathcal{O}^{\RM2}_i(\boldsymbol{q}, \boldsymbol{p}_1) g_i^{\text{NR}}(\boldsymbol{q}, \boldsymbol{p}_1),
\end{equation}
where the coefficients of the NR potential are given by
\begin{equation}\label{eq.ginr1}
\begin{aligned}
g_1^{\textrm{NR}}=&-g_{\mathrm{S}}^a g_{\mathrm{S}}^{b}\Big(1-\frac{4\boldsymbol{p}_1^2+\boldsymbol{q}^2+4\boldsymbol{q} \cdot \boldsymbol{p}_1}{8 m_{a}^2}-\frac{4\boldsymbol{p}_1^2+\boldsymbol{q}^2+4\boldsymbol{q} \cdot \boldsymbol{p}_1}{8m_{b}^2}\Big),\\
g_2^{\textrm{NR}}=&g_{\mathrm{S}}^a g_{\mathrm{S}}^{b} \mu ^2 \Big(\frac{1}{4 m_{a}^2}-\frac{4\boldsymbol{p}_1^2+\boldsymbol{q}^2+4\boldsymbol{q} \cdot \boldsymbol{p}_1}{32m_{a}^2m_{b}^2}-\frac{6 \boldsymbol{p}_1^2+3 \boldsymbol{q}^2+6 \boldsymbol{q} \cdot \boldsymbol{p}_1}{32 m_{a}^4}\Big),\\
g_3^{\textrm{NR}}=&g_{\mathrm{S}}^a g_{\mathrm{S}}^{b} \mu ^2 \Big(\frac{1}{4m_{b}^2}-\frac{4\boldsymbol{p}_1^2+\boldsymbol{q}^2+4\boldsymbol{q} \cdot \boldsymbol{p}_1}{32 m_{a}^2m_{b}^2}-\frac{6 \boldsymbol{p}_1^2+3 \boldsymbol{q}^2+6 \boldsymbol{q} \cdot \boldsymbol{p}_1}{32m_{b}^4}\Big),\\
g_4^{\textrm{NR}}=&g_{\mathrm{P}}^{a} g_{\mathrm{S}}^{b} \mu  \Big(\frac{1}{2 m_{a}}-\frac{4\boldsymbol{p}_1^2+\boldsymbol{q}^2+4\boldsymbol{q} \cdot \boldsymbol{p}_1}{16 m_{a}m_{b}^2}-\frac{4\boldsymbol{p}_1^2+3 \boldsymbol{q}^2+6 \boldsymbol{q} \cdot \boldsymbol{p}_1}{16 m_{a}^3}\Big),\\
g_5^{\textrm{NR}}=&- g_{\mathrm{S}}^ag_{\mathrm{P}}^{b} \mu  \Big(\frac{1}{2m_{b}}-\frac{4\boldsymbol{p}_1^2+\boldsymbol{q}^2+4\boldsymbol{q} \cdot \boldsymbol{p}_1}{16 m_{a}^2m_{b}}-\frac{4\boldsymbol{p}_1^2+3 \boldsymbol{q}^2+6 \boldsymbol{q} \cdot \boldsymbol{p}_1}{16m_{b}^3}\Big),\\
g_6^{\textrm{NR}}=&-g_{\mathrm{P}}^{a} g_{\mathrm{S}}^{b} \mu  \frac{\boldsymbol{q}^2+2\boldsymbol{q} \cdot \boldsymbol{p}_1}{8 m_{a}^3},\\
g_7^{\textrm{NR}}=& g_{\mathrm{S}}^ag_{\mathrm{P}}^{b} \mu  \frac{\boldsymbol{q}^2+2\boldsymbol{q} \cdot \boldsymbol{p}_1}{8m_{b}^3},\\
g_8^{\textrm{NR}}=&0,
\end{aligned}
\end{equation}
and
\begin{equation}\label{eq.ginr2}
\begin{aligned}
g_9^{\textrm{NR}}=&-g_{\mathrm{P}}^{a} g_{\mathrm{P}}^{b} \frac{\mu ^2}{4 m_{a}m_{b}} \Big(1-\frac{4\boldsymbol{p}_1^2+3 \boldsymbol{q}^2+6 \boldsymbol{q} \cdot \boldsymbol{p}_1}{8 m_{a}^2}-\frac{4\boldsymbol{p}_1^2+3 \boldsymbol{q}^2+6 \boldsymbol{q} \cdot \boldsymbol{p}_1}{8 m_{b}^2}\Big)-g_{\mathrm{S}}^a g_{\mathrm{S}}^{b} \mu ^2\frac{ \boldsymbol{p}_1^2}{16 m_{a}^2m_{b}^2},\\
g_{10}^{\textrm{NR}}=&g_{\mathrm{S}}^a g_{\mathrm{S}}^{b}  \mu ^2\frac{\boldsymbol{q} \cdot \boldsymbol{p}_1}{16 m_{a}^2m_{b}^2}+g_{\mathrm{P}}^{a} g_{\mathrm{P}}^{b} \mu ^2 \frac{\boldsymbol{q}^2+2\boldsymbol{q} \cdot \boldsymbol{p}_1}{16 m_{a}^3m_{b}},\\
g_{11}^{\textrm{NR}}=&g_{\mathrm{S}}^a g_{\mathrm{S}}^{b} \mu ^2\frac{ \boldsymbol{q} \cdot \boldsymbol{p}_1}{16 m_{a}^2m_{b}^2}+g_{\mathrm{P}}^{a} g_{\mathrm{P}}^{b}\mu ^2 \frac{\boldsymbol{q}^2+2\boldsymbol{q} \cdot \boldsymbol{p}_1}{16 m_{a}m_{b}^3},\\
g_{12}^{\textrm{NR}}=&-g_{\mathrm{S}}^a g_{\mathrm{S}}^{b} \mu^2 \frac{\boldsymbol{q}^2}{16 m_{a}^2m_{b}^2},\\
g_{13}^{\textrm{NR}}=&- g_{\mathrm{S}}^{a} g_{\mathrm{P}}^{b}\mu ^3 \Big(\frac{1}{8 m_{a}^2 m_{b}}-\frac{6 \boldsymbol{p}_1^2+3 \boldsymbol{q}^2+6 \boldsymbol{q} \cdot \boldsymbol{p}_1}{64 m_{a}^4 m_{b}}-\frac{4\boldsymbol{p}_1^2+3 \boldsymbol{q}^2+6 \boldsymbol{q} \cdot \boldsymbol{p}_1}{64 m_{a}^2 m_{b}^3}\Big)\\
&+g_{\mathrm{P}}^{a} g_{\mathrm{S}}^{b}\mu^3 \Big(\frac{1}{8 m_{a} m_{b}^2}-\frac{4\boldsymbol{p}_1^2+3 \boldsymbol{q}^2+6 \boldsymbol{q} \cdot \boldsymbol{p}_1}{64 m_{a}^3 m_{b}^2}-\frac{6 \boldsymbol{p}_1^2+3 \boldsymbol{q}^2+6 \boldsymbol{q} \cdot \boldsymbol{p}_1}{64 m_{a} m_{b}^4}\Big),\\
g_{14}^{\textrm{NR}}=&g_{\mathrm{S}}^{a}g_{\mathrm{P}}^{b}  \mu ^3\frac{\boldsymbol{q}^2+2\boldsymbol{q} \cdot \boldsymbol{p}_1}{32 m_{a}^2 m_{b}^3} -g_{\mathrm{P}}^{a} g_{\mathrm{S}}^{b} \mu ^3\frac{\boldsymbol{q}^2+2\boldsymbol{q} \cdot \boldsymbol{p}_1}{32 m_{a}^3 m_{b}^2},\\
g_{15}^{\textrm{NR}}=&\frac{g_{\mathrm{P}}^{a} g_{\mathrm{S}}^{b} \mu  \boldsymbol{q} \cdot \boldsymbol{p}_1}{8 m_{a} m_{b}^2},\\
g_{16}^{\textrm{NR}}=&-\frac{g_{\mathrm{P}}^{a} g_{\mathrm{S}}^{b} \mu  \boldsymbol{q}^2}{8 m_{a} m_{b}^2}.
\end{aligned}
\end{equation}

We would like to mention that the on-shell conditions are not applied in the above calculation, namely the off-shell effects are kept in the results of Eqs.~\eqref{eq.ginr1} and \eqref{eq.ginr2}. By imposing the on-shell condition, one will get
\begin{equation}
\begin{aligned}
\boldsymbol{q} \cdot \boldsymbol{p}_1=-\frac{1}{2}\boldsymbol{q}^2,
\end{aligned}
\end{equation}
and the potential can then take a different form when higher-order contributions are considered.

\section{Non-relativistic potential in the coordinate space}\label{sec.coordinate}

Although it is straightforward to calculate the relativistic potential in the momentum space, such as those in Eqs.~\eqref{eq.gir1} and \eqref{eq.gir2}, there is still no viable way to get its counterpart in the coordinate space. In contrast, the NR potential in coordinate space can be derived by performing the standard Fourier transformation. For the LO NR potential, the effects of the long-range part have been discussed in
Refs.~\cite{Dobrescu:2006au,Fadeev:2018rfl,Fadeev:2022wzg}, and the effects of the short-range part are analyzed in Refs.~\cite{Fadeev:2018rfl,Fadeev:2022wzg}.
To make a direct comparison with the potential in Ref.~\cite{Dobrescu:2006au}, we will first revisit the long-range part of $\overline{V}^{\RM1,\text{NR}}$ as discussed in the former reference. As an improvement, the higher-order contributions from the full relativistic and off-shell effects will be taken into account. More importantly, we will also perform Fourier transformation of the potentials in the new basis introduced in Sec.~\ref{sec.newopm} and derive their expressions in the coordinate space by simultaneously keeping the short-range and long-range parts.

\subsection{Long-range potential in coordinate space for the hybrid representation}

In Ref.~\cite{Dobrescu:2006au}, the variable $\boldsymbol{P}/m_a$ in the long-range potential part of $\overline{V}^{\RM1,\text{NR}}$~\eqref{eq.vinrm} is interpreted as the classical velocity, which remains unaffected by the Fourier transformation with respect to $\boldsymbol{q}$. Meanwhile,
the former reference only focuses on the long-range potential between two macroscopic objects, in which the short-range parts are not so relevant.

It is mentioned that most of the $\boldsymbol{q}^2$ terms in Eq.~\eqref{eq.vinrmf1}, except the one in $f_2$ (otherwise it will vanish), will be neglected when performing the Fourier transformation in the study of the hybrid representation. This is a valid approximation for a large class of models, since after the Fourier transformation $\boldsymbol{q}^2$ gives a factor of $m_0^2/m_a^2$, which is strongly suppressed due to the tiny mass $m_0$ of the exchanged spin-0 particle. Furthermore, since $\boldsymbol{P}/m_a$ is considered to be a classic velocity, instead of an operator, the spin- and $\boldsymbol{r}$-dependent structures for the on-shell parts of the potentials in coordinate space will remain the same with the LO ones as given in Ref.~\cite{Dobrescu:2006au}. The relativistic corrections manifest in some global factors related to the velocity squared. While, for the off-shell parts, different structures with spin and $\boldsymbol{r}$ dependences will appear.

The explicit expressions in the coordinate space can be obtained via the Fourier transformation
\begin{equation}
V^{\RM1,\text{NR}}(\boldsymbol{r},\boldsymbol{P})\equiv \int \frac{d^3\boldsymbol{q}}{(2\pi)^3}\overline{V}^{\RM1,\text{NR}}(\boldsymbol{q},\boldsymbol{P})e^{i\boldsymbol{q}\cdot\boldsymbol{r}}\equiv \sum V^{\RM1,\text{NR}}_{i}\equiv \sum V^{\RM1,\text{NR}}_{i,\textrm{on}}+ V^{\RM1,\text{NR}}_{i,\textrm{off}}.
\end{equation}
Utilizing the properties of Fourier transformation delineated in Appendix~\ref{sec.fourier}, one can match from the corresponding potentials of Eq.~\eqref{eq.vinrm} in momentum space to the forms in the coordinate space, and the terms with relativistic corrections that survive in the on-shell condition are found to be
\begin{equation}\label{eq.vinrc1}
\begin{aligned}
V_{1,\textrm{on}}^{\RM1,\text{NR}}=& -g_{\mathrm{S}}^{a}g_{\mathrm{S}}^{b}\Big[1-\frac{\boldsymbol{P}^2}{2m_{a}^2}-\frac{\boldsymbol{P}^2}{2m_{b}^2} \Big]V_0,\\
V_{2,\textrm{on}}^{\RM1,\text{NR}}=& -\boldsymbol{\sigma}^{a} \cdot \boldsymbol{\sigma}^{b}g_{\mathrm{S}}^{a}g_{\mathrm{S}}^{b} \frac{m_0^2\boldsymbol{P}^2}{16m_{a}^2m_{b}^2r}V_0,\\
V_{3,\textrm{on}}^{\RM1,\text{NR}}=&\Big[\boldsymbol{\sigma}^{a} \cdot \boldsymbol{\sigma}^{b}C_1-(\boldsymbol{\sigma}^{a} \cdot \boldsymbol{\bar{r}})(\boldsymbol{\sigma}^{b} \cdot \boldsymbol{\bar{r}})C_2\Big]\\
&\times\Big[-g_{\mathrm{S}}^{a} g_{\mathrm{S}}^{b}\frac{ \boldsymbol{P}^2}{16 m_{b}^2}+g_{\mathrm{P}}^{a} g_{\mathrm{P}}^{b} \Big(-\frac{m_{a}}{4 m_{b}}+\frac{ m_{a}\boldsymbol{P}^2}{8 m_{b}^3}+\frac{\boldsymbol{P}^2}{8 m_{a} m_{b}}\Big)\Big]\frac{1}{m_a^2r^2}V_0,\\
V_{4,\textrm{on}}^{\RM1,\text{NR}}=&(\boldsymbol{\sigma}^{a} + \boldsymbol{\sigma}^{b}) \cdot(\boldsymbol{P} \times \boldsymbol{\bar{r}})g_{\mathrm{S}}^{a} g_{\mathrm{S}}^{b} \Big(\frac{1}{4}+\frac{m_{a}^2}{4 m_{b}^2}-\frac{3 m_{a}^2 \boldsymbol{P}^2}{16 m_{b}^4}-\frac{\boldsymbol{P}^2}{4 m_{b}^2}-\frac{3 \boldsymbol{P}^2}{16 m_{a}^2}\Big)\frac{C_1}{2 m_a^2 r}V_0,\\
V_{5,\textrm{on}}^{\RM1,\text{NR}}=&-(\boldsymbol{\sigma}^{a} - \boldsymbol{\sigma}^{b}) \cdot(\boldsymbol{P} \times \boldsymbol{\bar{r}}) g_{\mathrm{S}}^{a} g_{\mathrm{S}}^{b} \Big(\frac{1}{4}-\frac{m_{a}^2}{4 m_{b}^2}+\frac{3 m_{a}^2 \boldsymbol{P}^2}{16 m_{b}^4}-\frac{3 \boldsymbol{P}^2}{16 m_{a}^2}\Big)\frac{C_1}{2 m_a^2 r}V_0,\\
V_{8,\textrm{on}}^{\RM1,\text{NR}}=& (\boldsymbol{\sigma}^{a} \cdot\boldsymbol{P}) (\boldsymbol{\sigma}^{b}\cdot \boldsymbol{P})g_{\mathrm{S}}^{a}g_{\mathrm{S}}^{b}\frac{m_0^2}{16m_{a}^2m_{b}^2r}V_0,\\
V_{9,\textrm{on}}^{\RM1,\text{NR}}=&(\boldsymbol{\sigma}^{a} + \boldsymbol{\sigma}^{b}) \cdot \boldsymbol{\bar{r}} \Big[g_{\mathrm{S}}^{a}g_{\mathrm{P}}^{b} \Big(\frac{m_{a}}{2 m_{b}}-\frac{ m_{a}\boldsymbol{P}^2}{4 m_{b}^3}-\frac{\boldsymbol{P}^2}{4 m_{a} m_{b}}\Big)-g_{\mathrm{P}}^{a} g_{\mathrm{S}}^{b}\Big(\frac{1}{2}-\frac{\boldsymbol{P}^2}{4 m_{a}^2}-\frac{\boldsymbol{P}^2}{4 m_{b}^2}\Big)\Big]\frac{C_1}{2 m_a r}V_0 ,\\
V_{10,\textrm{on}}^{\RM1,\text{NR}}=&(\boldsymbol{\sigma}^{a} - \boldsymbol{\sigma}^{b}) \cdot \boldsymbol{\bar{r}} \Big[g_{\mathrm{S}}^{a}g_{\mathrm{P}}^{b} \Big(\frac{m_{a}}{2 m_{b}}-\frac{ m_{a}\boldsymbol{P}^2}{4 m_{b}^3}-\frac{\boldsymbol{P}^2}{4 m_{a} m_{b}}\Big)+g_{\mathrm{P}}^{a} g_{\mathrm{S}}^{b}\Big(\frac{1}{2}-\frac{\boldsymbol{P}^2}{4 m_{a}^2}-\frac{\boldsymbol{P}^2}{4 m_{b}^2}\Big)\Big]\frac{C_1}{2 m_a r}V_0 ,\\
V_{14,\textrm{on}}^{\RM1,\text{NR}}=&-(\boldsymbol{\sigma}^{a}\times \boldsymbol{\sigma}^{b})\cdot \boldsymbol{P}\Big(g_{\mathrm{S}}^{a}g_{\mathrm{P}}^{b}m_{b}+g_{\mathrm{P}}^{a}g_{\mathrm{S}}^{b}m_a\Big)\frac{m_0^2}{16m_{a}^2m_{b}^2r}V_0,\\
V_{15,\textrm{on}}^{\RM1,\text{NR}}=&-\Big[\boldsymbol{\sigma}^{a} \cdot(\boldsymbol{P} \times \boldsymbol{\bar{r}})\boldsymbol{\sigma}^{b} \cdot \boldsymbol{\bar{r}}+\boldsymbol{\sigma}^{a} \cdot \boldsymbol{\bar{r}}\boldsymbol{\sigma}^{b} \cdot(\boldsymbol{P} \times \boldsymbol{\bar{r}})\Big]\Big(g_{\mathrm{S}}^{a} g_{\mathrm{P}}^{b}\frac{ m_{a}}{8 m_{b}}-g_{\mathrm{P}}^{a} g_{\mathrm{S}}^{b}\frac{ m_{a}^2}{8 m_{b}^2}\Big)\frac{1}{2 m_a^3 r^2}C_2V_0,
\end{aligned}
\end{equation}
and the terms contributed by the off-shell effects are given by
\begin{equation}\label{eq.vinrc2}
\begin{aligned}
V_{1,\textrm{off}}^{\RM1,\text{NR}}=& \frac{ 5 g_{\mathrm{S}}^{a}g_{\mathrm{S}}^{b}\left(m_{a}^4+m_{b}^4\right) }{32 m_{a}^4 m_{b}^4 r^2}
[C_1 \boldsymbol{P}^2-C_2 (\boldsymbol{P} \cdot \boldsymbol{\bar{r}})^2]V_0,\\
V_{2,\textrm{off}}^{\RM1,\text{NR}}=& -\boldsymbol{\sigma}^{a}\cdot\boldsymbol{\sigma}^{b} \frac{g_{\mathrm{S}}^{a}g_{\mathrm{S}}^{b}}{16 m_{a}^2 m_{b}^2 r^2}[C_1 \boldsymbol{P}^2-C_2 (\boldsymbol{P} \cdot \boldsymbol{\bar{r}})^2]V_0,\\
V_{6,\textrm{off}}^{\RM1,\text{NR}}=&[2C_1\boldsymbol{\sigma}^{a}\cdot \boldsymbol{P} \boldsymbol{\sigma} \cdot \boldsymbol{P}-C_2\boldsymbol{P} \cdot \boldsymbol{\bar{r}}(\boldsymbol{\sigma}^{a}\cdot \boldsymbol{\bar{r}} \boldsymbol{\sigma}^{b} \cdot \boldsymbol{P}+\boldsymbol{\sigma}^{a}\cdot \boldsymbol{P} \boldsymbol{\sigma}^{b} \cdot \boldsymbol{\bar{r}}) ] \frac{g_{\mathrm{S}}^{a}g_{\mathrm{S}}^{b}m_{a} m_{b}+g_{\mathrm{P}}^{a}g_{\mathrm{P}}^{b}\left(m_{a}^2+m_{b}^2\right)}{16 m_{a}^3 m_{b}^3 r^2}V_0,\\
V_{7,\textrm{off}}^{\RM1,\text{NR}}=&(\boldsymbol{\sigma}^{a} \cdot \boldsymbol{P} \boldsymbol{\sigma}^{b} \cdot \boldsymbol{\bar{r}}-\boldsymbol{\sigma}^{a} \cdot \boldsymbol{\bar{r}}  \boldsymbol{\sigma}^{b} \cdot \boldsymbol{P})\frac{g_{\mathrm{P}}^{a}g_{\mathrm{P}}^{b}\left(m_{a}^2-m_{b}^2\right) }{16 m_{a}^3 m_{b}^3 r^2} C_2 \boldsymbol{P} \cdot \boldsymbol{\bar{r}}V_0 ,\\
V_{11,\textrm{off}}^{\RM1,\text{NR}}=&(\boldsymbol{\sigma}^{a} \cdot \boldsymbol{P}+\boldsymbol{\sigma}^{b} \cdot \boldsymbol{P})\frac{-g_{\mathrm{S}}^{a}g_{\mathrm{P}}^{b}m_{a}^3 +g_{\mathrm{P}}^{a}g_{\mathrm{S}}^{b}m_{b}^3}{8 m_{a}^3 m_{b}^3 r} C_1 \boldsymbol{P} \cdot \boldsymbol{\bar{r}}V_0 ,\\
V_{12,\textrm{off}}^{\RM1,\text{NR}}=&(\boldsymbol{\sigma}^{a} \cdot \boldsymbol{P}-\boldsymbol{\sigma}^{b} \cdot \boldsymbol{P})\frac{g_{\mathrm{S}}^{a}g_{\mathrm{P}}^{b}m_{a}^3 +g_{\mathrm{P}}^{a}g_{\mathrm{S}}^{b}m_{b}^3 }{8 m_{a}^3 m_{b}^3 r}C_1 \boldsymbol{P} \cdot \boldsymbol{\bar{r}}V_0,\\
V_{13,\textrm{off}}^{\RM1,\text{NR}}=&[C_2\boldsymbol{P} \cdot \boldsymbol{\bar{r}}(\boldsymbol{\sigma}^{a}\times \boldsymbol{\sigma}^{b}) \cdot \boldsymbol{\bar{r}}- C_1 (\boldsymbol{\sigma}^{a}\times \boldsymbol{\sigma}^{b}) \cdot \boldsymbol{P} ] \frac{g_{\mathrm{S}}^{a}g_{\mathrm{P}}^{b}m_{b}+g_{\mathrm{P}}^{a}g_{\mathrm{S}}^{b}m_{a}}{16 m_{a}^2 m_{b}^2 r^2}V_0,
\end{aligned}
\end{equation}
with $r\equiv|\boldsymbol{r}|,\boldsymbol{\bar{r}}=\boldsymbol{r}/r$ and
\begin{equation}
\begin{aligned}
V_0\equiv&\frac{1}{4\pi r}e^{-m_0r},\\
C_1=&m_0r+1,\\
C_2=&m_0^2 r^2+3 m_0 r+3.
\label{eq:V0C1C2}
\end{aligned}
\end{equation}

By neglecting the higher-order contributions featured by the $\boldsymbol{P}^2$ and $\boldsymbol{P} \cdot \boldsymbol{\bar{r}}$ terms in the aforementioned expressions, we can exactly recover the results presented in Eqs.~(3.6, 3.8, 4.6, 4.7) of Ref.~\cite{Dobrescu:2006au}, with the exception of a global factor of $\frac{1}{2}$ in $V_{15}$ that already exists in the momentum space. It is noticed that eight terms, $V_{2,6,7,8,11,12,13,14}^{\RM1}$, which are zero at LO~\cite{Dobrescu:2006au}, receive non-vanishing higher-order contributions from the relativistic and off-shell effects. It is also interesting to point out that the NLO and NNLO corrections are absent for $V_{15}$, though it receives contribution at LO. In contrast, contributions that depend on the square of the velocity emerge for the other potentials. These new contributions may provide further insights for laboratory experiments, enabling more comprehensive analyses.

\subsection{NR potential in coordinate space for the new operator basis}

In the study of Ref.~\cite{Dobrescu:2006au}, the variable $\boldsymbol{P}/m_a$ is simply interpreted as the velocity.
% According to the quantum theory, the velocity is actually given by the expectation value of the momentum operator, i.e., $\langle p \rangle/m$.
Strictly speaking, for the Hamiltonian of a two-particle system in coordinate space, $\boldsymbol{\hat{p}}_1$ represents the momentum operator $-i\frac{\partial}{\partial \boldsymbol{r}}$, with $\boldsymbol{r}$ the relative position vector pointing from particle 1 to particle 2, and $\langle \hat{\boldsymbol{p}}_1 \rangle/m_1$ corresponds to the velocity of particle 1. Furthermore, in the Fourier transformation, the operator $\boldsymbol{\hat{p}}_1$ in potentials in the coordinate space should be placed at the far right of the expressions \cite{Berestetskii:1982qgu}. Since $\boldsymbol{P}$ is a mixture  of $\boldsymbol{p}_1$ and $\boldsymbol{q}$, the operator $\boldsymbol{\hat{P}}$ can not be positioned freely in the coordinate space. Therefore, the optimal approach is to use $\boldsymbol{p}_1$ and $\boldsymbol{q}$ as variables, instead of $\boldsymbol{q}$ and $\boldsymbol{P}$ as proposed in Ref.~\cite{Dobrescu:2006au}, to express the potential in momentum space and then transfer them to coordinate space. When discussing both the short-range and long-range potentials for the atomic systems, the counterpart of the terms with $\boldsymbol{q}\cdot\boldsymbol{p}_1$ in momentum space will be retained in our calculation when performing the Fourier transformation. The potential in the coordinate space is written as
\begin{equation}
V(\boldsymbol{r},\boldsymbol{\hat{p}}_1)\rightarrow V^{\RM3,\text{NR}} \equiv \sum_{i=1}^{16} V^{\RM3,\text{NR}}_{i}\equiv \sum_{i=1}^{16} \mathcal{O}^{\RM3}_i(\boldsymbol{r}, \boldsymbol{\hat{p}}_1) h_i^{\text{NR}}(\boldsymbol{r}, \boldsymbol{\hat{p}}_1),
\label{eq:potential-3}
\end{equation}
where the operator basis of $\mathcal{O}^{\RM3}_i(\boldsymbol{r}, \boldsymbol{\hat{p}}_1)$ has one-to-one correspondence with $\mathcal{O}^{\RM2}_i$ in Eqs.~\eqref{eq.oii1} and \eqref{eq.oii2}. The spin-independent and one-spin-dependent operators are
\begin{equation}
\begin{aligned}
\mathcal{O}^{\RM3}_{1} &\equiv 1,  \\
\mathcal{O}^{\RM3}_{2} &\equiv \frac{1}{\mu } \boldsymbol{\sigma}^{a}\cdot \boldsymbol{\hat{L}},
&\mathcal{O}^{\RM3}_{3} &\equiv \frac{1}{\mu }\boldsymbol{\sigma}^{b}\cdot \boldsymbol{\hat{L}}, \\
\mathcal{O}^{\RM3}_{4} &\equiv \boldsymbol{\sigma}^{a}\cdot \boldsymbol{\bar{r}},
&\mathcal{O}^{\RM3}_{5} &\equiv \boldsymbol{\sigma}^{b}\cdot \boldsymbol{\bar{r}},\\
\mathcal{O}^{\RM3}_{6} &\equiv \frac{i}{\mu} \boldsymbol{\sigma}^{a}\cdot \boldsymbol{\hat{p}}_1,
&\mathcal{O}^{\RM3}_{7} &\equiv \frac{i}{\mu}\boldsymbol{\sigma}^{b}\cdot \boldsymbol{\hat{p}}_1,
\end{aligned}
\label{eq:operator-3-1}
\end{equation}
and the two-spin-dependent operators are
\begin{equation}
\begin{aligned}
\mathcal{O}^{\RM3}_{8} &\equiv \boldsymbol{\sigma}^{a}\cdot \boldsymbol{\sigma}^{b},\\
\mathcal{O}^{\RM3}_{9} &\equiv \boldsymbol{\sigma}^{a}\cdot \boldsymbol{\bar{r}} \boldsymbol{\sigma}^{b}\cdot \boldsymbol{\bar{r}},
&\mathcal{O}^{\RM3}_{10}&\equiv \frac{i}{\mu}  \boldsymbol{\sigma}^{b}\cdot \boldsymbol{\bar{r}}\boldsymbol{\sigma}^{a}\cdot \boldsymbol{\hat{p}}_1,\\
\mathcal{O}^{\RM3}_{11}&\equiv \frac{i}{\mu}\boldsymbol{\sigma}^{a}\cdot \boldsymbol{\bar{r}} \boldsymbol{\sigma}^{b}\cdot \boldsymbol{\hat{p}}_1,
&\mathcal{O}^{\RM3}_{12}&\equiv \frac{1}{\mu^2}\boldsymbol{\sigma}^{a}\cdot \boldsymbol{\hat{p}}_1 \boldsymbol{\sigma}^{b}\cdot \boldsymbol{\hat{p}}_1, \\
\mathcal{O}^{\RM3}_{13}&\equiv \frac{1}{\mu}\boldsymbol{\sigma}^{a}\cdot \boldsymbol{\hat{L}}\boldsymbol{\sigma}^{b}\cdot \boldsymbol{\bar{r}},
&\mathcal{O}^{\RM3}_{14}&\equiv \frac{i}{\mu^2 }\boldsymbol{\sigma}^{a}\cdot \boldsymbol{\hat{L}} \boldsymbol{\sigma}^{b}\cdot \boldsymbol{\hat{p}}_1, \\
\mathcal{O}^{\RM3}_{15}&\equiv i(\boldsymbol{\sigma}^{a}\times \boldsymbol{\sigma}^{b}) \cdot \boldsymbol{\bar{r}},
&\mathcal{O}^{\RM3}_{16}&\equiv\frac{1}{\mu}(\boldsymbol{\sigma}^{a}\times \boldsymbol{\sigma}^{b}) \cdot \boldsymbol{\hat{p}}_1\,,
\end{aligned}
\label{eq:operator-3-2}
\end{equation}
with $\boldsymbol{\hat{L}}\equiv \boldsymbol{r}\times\boldsymbol{\hat{p}}_1$. We clarify that the operator $\boldsymbol{\hat{p}}_1$ in the potential is understood to operate solely on the wave function and does not act on $r$ or $\boldsymbol{\bar{r}}$ in the potential. The coefficients $h_{i}^{\textrm{NR}}(\boldsymbol{r},\boldsymbol{\hat{p}}_1)$ in Eq.~\eqref{eq:potential-3} can be decomposed as
\begin{equation}
\begin{aligned}
h_{i}^{\textrm{NR}}(\boldsymbol{r},\boldsymbol{\hat{p}}_1) \equiv h_{i,1}^{\textrm{NR}}+h_{i,2}^{\textrm{NR}}\boldsymbol{\hat{p}}_1^2+ih_{i,3}^{\textrm{NR}}\boldsymbol{\bar{r}}\cdot\boldsymbol{\hat{p}}_1
+\pi\delta^3(\boldsymbol{\bar{r}})(h_{i,4}^{\textrm{NR}}+h_{i,5}^{\textrm{NR}}\boldsymbol{\hat{p}}_1^2+ih_{i,6}^{\textrm{NR}}\boldsymbol{\bar{r}}\cdot \boldsymbol{\hat{p}}_1),
\end{aligned}
\label{eq:expression-hi}
\end{equation}
where $h_{i,j}^{\textrm{NR}}$ are only dependent on $r$, whose expressions are presented in Appendix~\ref{sec.hij}.

Employing Eqs.~(\ref{eq:potential-3}-\ref{eq:expression-hi}) and the expressions for $h_{i,j}^{\textrm{NR}}$ in Appendix~\ref{sec.hij}, one can readily derive the potential in the forms utilized in Refs.~\cite{Fadeev:2018rfl,Fadeev:2022wzg}. As an illustrative example, upon expanding the coefficients $h_{i,j}^{\textrm{NR}}$ in terms of a joint expansion of $1/m_a$ and $1/m_{b}$ to second order, one obtains
\begin{equation}
\begin{aligned}
V_{pp}=&-\frac{g_{\mathrm{P}}^{a} g_{\mathrm{P}}^{b}W}{4m_am_{b}r^3}\Big\{\mathcal{O}^{\RM3}_8[C_1-4\pi r^3\delta^3(\boldsymbol{r})]-\mathcal{O}^{\RM3}_9[C_2-16\pi r^3\delta^3(\boldsymbol{r})]\\
&~~~~~~~~~~~~~~~~~~-\mathcal{O}^{\RM3}_{12}\frac{m_a^2+m_{b}^2}{2(m_a+m_{b})^2}[C_1-4\pi r^3\delta^3(\boldsymbol{r})]\Big\} \\
=&-\frac{g_{\mathrm{P}}^{a} g_{\mathrm{P}}^{b}W}{4m_am_{b}r^3}\Big\{\mathcal{O}^{\RM3}_8[C_1+\frac{4\pi}{3}r^3\delta^3(\boldsymbol{r})]- \mathcal{O}^{\RM3}_9C_2-\mathcal{O}^{\RM3}_{12}\frac{m_a^2+m_{b}^2}{2(m_a+m_{b})^2}[C_1-4\pi r^3\delta^3(\boldsymbol{r})]\Big\},
\end{aligned}
\label{eq:potential-Vpp}
\end{equation}
wherein the relations in Eq.~(\ref{eq:propety-of-delta-function-2}) has been used. In Eq. (\ref{eq:potential-Vpp}), the first two terms are identical to those presented in Eq. (3) of Ref. \cite{Fadeev:2022wzg}, while the third term is new and corresponds to the NLO relativistic contribution.  Analogously, we further derive other types of potentials contributed by different combinations of $g_S^{(a,b)}$ and $g_P^{(a,b)}$ couplings:
\begin{equation}\label{eq.vssvspvps}
\begin{aligned}
V_{ss}=&-g_{\mathrm{S}}^{a} g_{\mathrm{S}}^{b}W\Big\{\mathcal{O}^{\RM3}_1\Big[\frac{1}{r}+\frac{(m_a^2+m_{b}^2)[-4iC_1 \boldsymbol{\bar{r}}\cdot \boldsymbol{\hat{p}}_1-4 r\boldsymbol{\hat{p}}_1^2-4 \pi  r^2 \delta^3(\boldsymbol{r})+m_{0}^2 r]}{8 m_a^2 m_{b}^2 r^2 }\Big]\\
&~~~~~~~~~~~~~~+ \mathcal{O}^{\RM3}_2\frac{\mu C_1}{4m_a^2r^2}+ \mathcal{O}^{\RM3}_3\frac{\mu C_1}{4m_{b}^2r^2}
+\mathcal{O}^{\RM3}_{12}\frac{\mu ^2 [C_1-C_2+12 \pi  r^3 \delta^3(\boldsymbol{r})]}{16 m_a^2 m_{b}^2 r^3}
\Big\},\\
V_{sp}=&g_{\mathrm{S}}^{a} g_{\mathrm{P}}^{b}W\Big\{ \mathcal{O}^{\RM3}_5  \frac{C_1}{2 m_{b} r^2}\\
&+\mathcal{O}^{\RM3}_7\Big[\frac{\mu[2iC_1 m_a^2 r \boldsymbol{\bar{r}}\cdot \boldsymbol{\hat{p}}_1+2C_1 m_{b}^2+(6C_1-C_2)m_a^2]}{8m_a^2 m_{b}^3 r^3}-\frac{\pi \mu (m_a^2+m_{b}^2)\delta^3(\boldsymbol{r})}{m_a^2 m_{b}^3}\Big]\\
& +\mathcal{O}^{\RM3}_{13}\frac{\mu[C_2-16\pi r^3 \delta^3(\boldsymbol{r})]}{8m_a^2 m_{b} r^3}-\mathcal{O}^{\RM3}_{16}\frac{\mu[C_1-4\pi r^3 \delta^3(\boldsymbol{r})]}{8m_a^2 m_{b} r^3} \Big\},\\
V_{ps}=&-g_{\mathrm{P}}^{a} g_{\mathrm{S}}^{b}W\Big\{\mathcal{O}^{\RM3}_4\frac{C_1}{2 m_a r^2}\\
&+\mathcal{O}^{\RM3}_6\Big[\frac{\mu [2iC_1 m_{b}^2 r \boldsymbol{\bar{r}}\cdot \boldsymbol{\hat{p}}_1+2C_1 m_a^2+(6C_1-C_2)m_{b}^2]}{8m_a^3 m_{b}^2 r^3}-\frac{\pi\mu(m_a^2+m_{b}^2)\delta^3(\boldsymbol{r}) }{m_a^3m_{b}^2}\Big]\\
&+\mathcal{O}^{\RM3}_{13}\frac{\mu [C_2-16 \pi r^3 \delta^3(\boldsymbol{r})]}{8m_a m_{b}^2 r^3}+\mathcal{O}^{\RM3}_{16}\frac{\mu[C_1-C_2+12 \pi r^3 \delta^3(\boldsymbol{r})]}{8m_a m_{b}^2 r^3}\Big\}.
\end{aligned}
\end{equation}

The newly calculated potentials in Eq.~\eqref{eq.vssvspvps} and the updated one in Eq.~\eqref{eq:potential-Vpp} are ready for use to conduct the laboratory experimental analyses in future.

\section{Summary}\label{sec.summary}

In this study, we carry out an in-depth calculation of the interaction potentials with spin-0 particle exchange between two spin-1/2 fermions. Our calculation is based on the framework of Bethe-Salpeter equation, in conjunction with scattering amplitude analysis. Relativistic contributions and off-shell effects are carefully taken into account. The potentials in both momentum- and coordinate-space representations are calculated.  The on-shell results at leading order in the non-relativistic expansion in our study coincide with those reported in Ref.~\cite{Dobrescu:2006au}, with the exception of a global factor $\frac{1}{2}$ in $V_{15}^{\RM1}$. It is found that several terms in the potential receive non-vanishing corrections from the relativistic and off-shell contributions.

Apart from the revised calculation along the line of Ref.~\cite{Dobrescu:2006au}, we propose a different operator basis to construct the interaction potential, in order to conveniently calculate the potential in the coordinate space. Both relativistic and off-shell effects are included to derive the interaction potentials with the new operator basis in the momentum and coordinate spaces. We hope that the new potential forms provided in this work can be useful to analyze laboratory experiments.

\section*{Acknowledgments}

This work is funded in part by the National Natural Science Foundations of China under Grants Nos.~12150013, 12075058 and 11975090, and the Science Foundation of Hebei Normal University with contract No.~L2023B09.

\section*{Appendix}
\subsection{Expressions for $h_{i,j}^{\textrm{NR}}$ }\label{sec.hij}
\setcounter{equation}{0}
\renewcommand\theequation{A.\arabic{equation}}
In this Appexdix, we list the expressions for $h_{i,j}^{\textrm{NR}}$~defined in Eq.~\eqref{eq:expression-hi}. The coefficients $h_{1,j}^{\textrm{NR}}$ take the form
\begin{equation}
\begin{aligned}
h_{1,1}^{\textrm{NR}}=&-g_{\mathrm{S}}^{a} g_{\mathrm{S}}^{b} \frac{m_0^2(m_{a}^2+m_{b}^2) +8 m_{a}^2 m_{b}^2}{8 m_{a}^2 m_{b}^2 r},\\
h_{1,2}^{\textrm{NR}}=&g_{\mathrm{S}}^{a} g_{\mathrm{S}}^{b}\frac{ m_{a}^2+m_{b}^2}{2 m_{a}^2 m_{b}^2 r},\\
h_{1,3}^{\textrm{NR}}=&g_{\mathrm{S}}^{a} g_{\mathrm{S}}^{b}\frac{(m_{a}^2+m_{b}^2) C_1}{2 m_{a}^2 m_{b}^2 r^2},\\
h_{1,4}^{\textrm{NR}}=&g_{\mathrm{S}}^{a} g_{\mathrm{S}}^{b}\frac{m_a^2+m_{b}^2}{2m_{a}^2m_{b}^2},\\
h_{1,5}^{\textrm{NR}}=&0,\\
h_{1,6}^{\textrm{NR}}=&0.
\end{aligned}
\end{equation}
The coefficients $h_{2,j}^{\textrm{NR}}$ are given by
\begin{equation}
\begin{aligned}
h_{2,1}^{\textrm{NR}}=&-g_{\mathrm{S}}^{a} g_{\mathrm{S}}^{b}\frac{ \mu [m_0^2(m_{a}^2+3 m_{b}^2) +8 m_{a}^2 m_{b}^2]  C_1}{32 m_{a}^4 m_{b}^2 r^2},\\
h_{2,2}^{\textrm{NR}}=&g_{\mathrm{S}}^{a} g_{\mathrm{S}}^{b} \frac{ \mu (2 m_{a}^2+3 m_{b}^2) C_1}{16 m_{a}^4 m_{b}^2 r^2},\\
h_{2,3}^{\textrm{NR}}=&g_{\mathrm{S}}^{a} g_{\mathrm{S}}^{b} \frac{ \mu (2 m_{a}^2+3 m_{b}^2) C_2}{16 m_{a}^4 m_{b}^2 r^3},\\
h_{2,4}^{\textrm{NR}}=&g_{\mathrm{S}}^{a} g_{\mathrm{S}}^{b} \frac{3 \mu (m_{a}^2+3 m_{b}^2) }{8 m_{a}^4 m_{b}^2 r},\\
h_{2,5}^{\textrm{NR}}=&0,\\
h_{2,6}^{\textrm{NR}}=&-g_{\mathrm{S}}^{a} g_{\mathrm{S}}^{b} \frac{\mu   (2 m_{a}^2+3 m_{b}^2) }{m_{a}^4 m_{b}^2}.
\end{aligned}
\end{equation}
The coefficients $h_{3,j}^{\textrm{NR}}$ are given by
\begin{equation}
\begin{aligned}
h_{3,1}^{\textrm{NR}}=&-g_{\mathrm{S}}^{a} g_{\mathrm{S}}^{b}\frac{ \mu [m_0^2(3 m_{a}^2+m_{b}^2) +8 m_{a}^2 m_{b}^2]  C_1}{32 m_{a}^2 m_{b}^4 r^2},\\
h_{3,2}^{\textrm{NR}}=&g_{\mathrm{S}}^{a} g_{\mathrm{S}}^{b}\frac{\mu (3 m_{a}^2+2 m_{b}^2)  C_1}{16 m_{a}^2 m_{b}^4 r^2},\\
h_{3,3}^{\textrm{NR}}=&g_{\mathrm{S}}^{a} g_{\mathrm{S}}^{b}\frac{ \mu (3 m_{a}^2+2 m_{b}^2)  C_2}{16 m_{a}^2 m_{b}^4 r^3},\\
h_{3,4}^{\textrm{NR}}=&g_{\mathrm{S}}^{a} g_{\mathrm{S}}^{b}\frac{3 \mu (3 m_{a}^2+m_{b}^2)  }{8 m_{a}^2 m_{b}^4 r},\\
h_{3,5}^{\textrm{NR}}=&0,\\
h_{3,6}^{\textrm{NR}}=&-g_{\mathrm{S}}^{a} g_{\mathrm{S}}^{b}\frac{ \mu  (3 m_{a}^2+2 m_{b}^2) }{m_{a}^2 m_{b}^4}.
\end{aligned}
\end{equation}
The coefficients $h_{4,j}^{\textrm{NR}}$ are given by
\begin{equation}
\begin{aligned}
h_{4,1}^{\textrm{NR}}=&-g_{\mathrm{P}}^{a} g_{\mathrm{S}}^{b} \frac{[m_0^2(m_{a}^2+3 m_{b}^2) +8 m_{a}^2 m_{b}^2] C_1}{16 m_{a}^3 m_{b}^2 r^2},\\
h_{4,2}^{\textrm{NR}}=&g_{\mathrm{P}}^{a} g_{\mathrm{S}}^{b} \frac{(m_{a}^2+m_{b}^2) C_1}{4 m_{a}^3 m_{b}^2 r^2},\\
h_{4,3}^{\textrm{NR}}=&g_{\mathrm{P}}^{a} g_{\mathrm{S}}^{b}\frac{ (2 m_{a}^2+3 m_{b}^2) C_2}{8 m_{a}^3 m_{b}^2 r^3},\\
h_{4,4}^{\textrm{NR}}=& g_{\mathrm{P}}^{a} g_{\mathrm{S}}^{b}\frac{3(m_{a}^2+3 m_{b}^2)  }{4 m_{a}^3 m_{b}^2 r},\\
h_{4,5}^{\textrm{NR}}=&0,\\
h_{4,6}^{\textrm{NR}}=&-g_{\mathrm{P}}^{a} g_{\mathrm{S}}^{b}\frac{2(2 m_{a}^2+3 m_{b}^2)  }{m_{a}^3 m_{b}^2}.
\end{aligned}
\end{equation}
The coefficients $h_{5,j}^{\textrm{NR}}$ are given by
\begin{equation}
\begin{aligned}
h_{5,1}^{\textrm{NR}}=&g_{\mathrm{S}}^{a}g_{\mathrm{P}}^{b}  \frac{[m_0^2(3 m_{a}^2+m_{b}^2) +8 m_{a}^2 m_{b}^2] C_1}{16 m_{a}^2 m_{b}^3 r^2},\\
h_{5,2}^{\textrm{NR}}=&- g_{\mathrm{S}}^{a}g_{\mathrm{P}}^{b} \frac{(m_{a}^2+m_{b}^2) C_1}{4 m_{a}^2 m_{b}^3 r^2},\\
h_{5,3}^{\textrm{NR}}=&-g_{\mathrm{S}}^{a}g_{\mathrm{P}}^{b} \frac{ (3 m_{a}^2+2 m_{b}^2) C_2}{8 m_{a}^2 m_{b}^3 r^3},\\
h_{5,4}^{\textrm{NR}}=&-g_{\mathrm{S}}^{a}g_{\mathrm{P}}^{b} \frac{3 (3 m_{a}^2+m_{b}^2) }{4 m_{a}^2 m_{b}^3 r},\\
h_{5,5}^{\textrm{NR}}=&0,\\
h_{5,6}^{\textrm{NR}}=& g_{\mathrm{S}}^{a}g_{\mathrm{P}}^{b} \frac{(4m_{b}^2+6m_{a}^2) }{m_{b}^3m_{a}^2}.
\end{aligned}
\end{equation}
The coefficients $h_{6,j}^{\textrm{NR}}$ are written as
\begin{equation}
\begin{aligned}
h_{6,1}^{\textrm{NR}}=&-g_{\mathrm{P}}^{a} g_{\mathrm{S}}^{b}\frac{ \mu  [2 m_{a}^2C_1 +m_{b}^2 (-m_0^2 r^2+3 m_0 r+3)]}{8 m_{a}^3 m_{b}^2 r^3},\\
h_{6,2}^{\textrm{NR}}=&0,\\
h_{6,3}^{\textrm{NR}}=&-g_{\mathrm{P}}^{a} g_{\mathrm{S}}^{b}\frac{ \mu  C_1}{4 m_{a}^3 r^2},\\
h_{6,4}^{\textrm{NR}}=&g_{\mathrm{P}}^{a} g_{\mathrm{S}}^{b}\frac{\mu (m_{a}^2+m_{b}^2)  }{m_{a}^3 m_{b}^2},\\
h_{6,5}^{\textrm{NR}}=&0,\\
h_{6,6}^{\textrm{NR}}=&0.
\end{aligned}
\end{equation}
The coefficients $h_{7,j}^{\textrm{NR}}$ are given by
\begin{equation}
\begin{aligned}
h_{7,1}^{\textrm{NR}}=&-g_{\mathrm{S}}^{a}g_{\mathrm{P}}^{b}  \frac{\mu  [m_{a}^2 (m_0^2 r^2-3 m_0 r-3)-2 m_{b}^2 C_1]}{8 m_{a}^2 m_{b}^3 r^3},\\
h_{7,2}^{\textrm{NR}}=&0,\\
h_{7,3}^{\textrm{NR}}=&g_{\mathrm{S}}^{a}g_{\mathrm{P}}^{b} \frac{ \mu  C_1}{4 m_{b}^3 r^2},\\
h_{7,4}^{\textrm{NR}}=&-g_{\mathrm{S}}^{a}g_{\mathrm{P}}^{b}  \frac{\mu (m_{a}^2+m_{b}^2) }{m_{a}^2 m_{b}^3},\\
h_{7,5}^{\textrm{NR}}=&0,\\
h_{7,6}^{\textrm{NR}}=&0.
\end{aligned}
\end{equation}
The coefficients $h_{8,j}^{\textrm{NR}}$ are given by
\begin{equation}
\begin{aligned}
h_{8,1}^{\textrm{NR}}=&-g_{\mathrm{P}}^{a} g_{\mathrm{P}}^{b}\frac{ (3 m_0^2(m_{a}^2+m_{b}^2) +8 m_{a}^2 m_{b}^2) C_1}{32 m_{a}^3 m_{b}^3 r^3},\\
h_{8,2}^{\textrm{NR}}=&-\frac{g_{\mathrm{S}}^{a} g_{\mathrm{S}}^{b} C_1}{16 m_{a}^2 m_{b}^2 r^3}+g_{\mathrm{P}}^{a} g_{\mathrm{P}}^{b}\frac{ (m_{a}^2+m_{b}^2) C_1}{8 m_{a}^3 m_{b}^3 r^3},\\
h_{8,3}^{\textrm{NR}}=& g_{\mathrm{P}}^{a} g_{\mathrm{P}}^{b}\frac{3 (m_{a}^2+m_{b}^2) C_2}{16 m_{a}^3 m_{b}^3 r^4},\\
h_{8,4}^{\textrm{NR}}=&g_{\mathrm{P}}^{a} g_{\mathrm{P}}^{b}\frac{ m_{a}^2(3 m_0^2 r^2+8 m_{b}^2 r^2+9) +3 m_{b}^2 (m_0^2 r^2+3)}{8 m_{a}^3 m_{b}^3 r^2},\\
h_{8,5}^{\textrm{NR}}=&g_{\mathrm{S}}^{a} g_{\mathrm{S}}^{b} \frac{1}{4 m_{a}^2 m_{b}^2}-g_{\mathrm{P}}^{a} g_{\mathrm{P}}^{b}\frac{ m_{a}^2+m_{b}^2}{2 m_{a}^3 m_{b}^3},\\
h_{8,6}^{\textrm{NR}}=&-g_{\mathrm{P}}^{a} g_{\mathrm{P}}^{b}\frac{3  (m_{a}^2+m_{b}^2)  }{m_{a}^3 m_{b}^3 r}.
\end{aligned}
\end{equation}
The coefficients $h_{9,j}^{\textrm{NR}}$ are given by
\begin{equation}
\begin{aligned}
h_{9,1}^{\textrm{NR}}=&g_{\mathrm{P}}^{a} g_{\mathrm{P}}^{b}\frac{[3 m_0^2 (m_{a}^2+m_{b}^2) +8 m_{a}^2 m_{b}^2] C_2}{32 m_{a}^3 m_{b}^3 r^3},\\
h_{9,2}^{\textrm{NR}}=&g_{\mathrm{S}}^{a} g_{\mathrm{S}}^{b}\frac{ C_2}{16 m_{a}^2 m_{b}^2 r^3}-g_{\mathrm{P}}^{a} g_{\mathrm{P}}^{b}\frac{ (m_{a}^2+m_{b}^2) C_2}{8 m_{a}^3 m_{b}^3 r^3},\\
h_{9,3}^{\textrm{NR}}=&- g_{\mathrm{P}}^{a} g_{\mathrm{P}}^{b} \frac{3(m_{a}^2+m_{b}^2) (m_0^3 r^3+6 m_0^2 r^2+15 m_0 r+15)}{16 m_{a}^3 m_{b}^3 r^4},\\
h_{9,4}^{\textrm{NR}}=&-g_{\mathrm{P}}^{a} g_{\mathrm{P}}^{b}\frac{   m_{a}^2(12 m_0^2 r^2+32 m_{b}^2 r^2+45)+3 m_{b}^2 (4 m_0^2 r^2+15)}{8 m_{a}^3 m_{b}^3 r^2},\\
h_{9,5}^{\textrm{NR}}=&-g_{\mathrm{S}}^{a} g_{\mathrm{S}}^{b}\frac{ 1 }{m_{a}^2 m_{b}^2}+g_{\mathrm{P}}^{a} g_{\mathrm{P}}^{b}\frac{2 (m_{a}^2+m_{b}^2)  }{m_{a}^3 m_{b}^3},\\
h_{9,6}^{\textrm{NR}}=& g_{\mathrm{P}}^{a} g_{\mathrm{P}}^{b} \frac{69  (m_{a}^2+m_{b}^2)}{4 m_{a}^3 m_{b}^3 r}.
\end{aligned}
\end{equation}
The coefficients $h_{10,j}^{\textrm{NR}}$ are given by
\begin{equation}
\begin{aligned}
h_{10,1}^{\textrm{NR}}=&g_{\mathrm{P}}^{a} g_{\mathrm{P}}^{b}\frac{ \mu [3 m_{a}^2C_2 +m_{b}^2 (-m_0^3 r^3+2 m_0^2 r^2+9 m_0 r+9)]}{16 m_{a}^3 m_{b}^3 r^4},\\
h_{10,2}^{\textrm{NR}}=&0,\\
h_{10,3}^{\textrm{NR}}=&g_{\mathrm{S}}^{a} g_{\mathrm{S}}^{b}\frac{ \mu  C_2}{16 m_{a}^2 m_{b}^2 r^3}+g_{\mathrm{P}}^{a} g_{\mathrm{P}}^{b}\frac{ \mu  C_2}{8 m_{a}^3 m_{b} r^3},\\
h_{10,4}^{\textrm{NR}}=&-g_{\mathrm{P}}^{a} g_{\mathrm{P}}^{b}\frac{3  \mu  (4 m_{a}^2+3 m_{b}^2) }{4 m_{a}^3 m_{b}^3 r},\\
h_{10,5}^{\textrm{NR}}=&0,\\
h_{10,6}^{\textrm{NR}}=&-g_{\mathrm{S}}^{a} g_{\mathrm{S}}^{b} \frac{ \mu }{m_{a}^2 m_{b}^2}- g_{\mathrm{P}}^{a} g_{\mathrm{P}}^{b}\frac{2  \mu }{m_{a}^3 m_{b}}.
\end{aligned}	
\end{equation}
The coefficients $h_{11,j}^{\textrm{NR}}$ are given by
\begin{equation}
\begin{aligned}
h_{11,1}^{\textrm{NR}}=&-g_{\mathrm{P}}^{a} g_{\mathrm{P}}^{b} \frac{\mu  [m_{a}^2 (m_0^3 r^3-2 m_0^2 r^2-9 m_0 r-9)-3 m_{b}^2 C_2]}{16 m_{a}^3 m_{b}^3 r^4},\\
h_{11,2}^{\textrm{NR}}=&0,\\
h_{11,3}^{\textrm{NR}}=&g_{\mathrm{S}}^{a} g_{\mathrm{S}}^{b}\frac{ \mu  C_2}{16 m_{a}^2 m_{b}^2 r^3}+g_{\mathrm{P}}^{a} g_{\mathrm{P}}^{b}\frac{ \mu  C_2}{8 m_{a} m_{b}^3 r^3},\\
h_{11,4}^{\textrm{NR}}=&- g_{\mathrm{P}}^{a} g_{\mathrm{P}}^{b}\frac{3\mu    (3 m_{a}^2+4 m_{b}^2) }{4 m_{a}^3 m_{b}^3 r},\\
h_{11,5}^{\textrm{NR}}=&0,\\
h_{11,6}^{\textrm{NR}}=&-g_{\mathrm{S}}^{a} g_{\mathrm{S}}^{b}\frac{  \mu }{m_{a}^2 m_{b}^2}- g_{\mathrm{P}}^{a} g_{\mathrm{P}}^{b} \frac{2  \mu }{m_{a} m_{b}^3}.
\end{aligned}
\end{equation}
The coefficients $h_{12,j}^{\textrm{NR}}$ are given by
\begin{equation}
\begin{aligned}
h_{12,1}^{\textrm{NR}}=&g_{\mathrm{S}}^{a} g_{\mathrm{S}}^{b}\frac{ \mu ^2(m_0^2 r^2+2 m_0 r+2) }{16 m_{a}^2 m_{b}^2 r^3}+g_{\mathrm{P}}^{a} g_{\mathrm{P}}^{b}\frac{ \mu ^2(m_{a}^2+m_{b}^2) C_1 }{8 m_{a}^3 m_{b}^3 r^3},\\
h_{12,2}^{\textrm{NR}}=&0,\\
h_{12,3}^{\textrm{NR}}=&0,\\
h_{12,4}^{\textrm{NR}}=&- g_{\mathrm{S}}^{a} g_{\mathrm{S}}^{b}\frac{3 \mu ^2}{4 m_{a}^2 m_{b}^2}-g_{\mathrm{P}}^{a} g_{\mathrm{P}}^{b}\frac{\mu ^2(m_{a}^2+m_{b}^2) }{2 m_{a}^3 m_{b}^3},\\
h_{12,5}^{\textrm{NR}}=&0,\\
h_{12,6}^{\textrm{NR}}=&0.
\end{aligned}
\end{equation}
The coefficients $h_{13,j}^{\textrm{NR}}$ are given by
\begin{equation}
\begin{aligned}
h_{13,1}^{\textrm{NR}}=&g_{\mathrm{S}}^{a}g_{\mathrm{P}}^{b} \frac{\mu  [3m_0^2(m_{a}^2+m_{b}^2) +8 m_{a}^2 m_{b}^2]  C_2}{64 m_{a}^4 m_{b}^3 r^3}\\&-g_{\mathrm{P}}^{a} g_{\mathrm{S}}^{b} \frac{\mu [3 m_0^2 (m_{a}^2+m_{b}^2)+8 m_{a}^2 m_{b}^2]  C_2}{64 m_{a}^3 m_{b}^4 r^3},\\
h_{13,2}^{\textrm{NR}}=&- g_{\mathrm{S}}^{a}g_{\mathrm{P}}^{b}\frac{  \mu (2 m_{a}^2+3 m_{b}^2) C_2}{32 m_{a}^4 m_{b}^3 r^3}+g_{\mathrm{P}}^{a} g_{\mathrm{S}}^{b}\frac{\mu  (3 m_{a}^2+2 m_{b}^2)  C_2}{32 m_{a}^3 m_{b}^4 r^3},\\
h_{13,3}^{\textrm{NR}}=& - g_{\mathrm{S}}^{a}g_{\mathrm{P}}^{b}\frac{3   \mu (m_{a}^2+m_{b}^2) (m_0^3 r^3+6 m_0^2 r^2+15 m_0 r+15)}{32 m_{a}^4 m_{b}^3 r^4}\\&+g_{\mathrm{P}}^{a} g_{\mathrm{S}}^{b}\frac{3\mu  (m_{a}^2+m_{b}^2)  (m_0^3 r^3+6 m_0^2 r^2+15 m_0 r+15)}{32 m_{a}^3 m_{b}^4 r^4},\\
h_{13,4}^{\textrm{NR}}=&-g_{\mathrm{S}}^{a}g_{\mathrm{P}}^{b} \frac{ \mu  [m_{a}^2(12 m_0^2 r^2+32 m_{b}^2 r^2+45) +3 m_{b}^2 (4 m_0^2 r^2+15)]}{16 m_{a}^4 m_{b}^3 r^2}\\
&+g_{\mathrm{P}}^{a} g_{\mathrm{S}}^{b}\frac{ \mu [m_{a}^2(12 m_0^2 r^2+32 m_{b}^2 r^2+45) +3 m_{b}^2 (4 m_0^2 r^2+15)]}{16 m_{a}^3 m_{b}^4 r^2},\\
h_{13,5}^{\textrm{NR}}=&g_{\mathrm{S}}^{a}g_{\mathrm{P}}^{b} \frac{ \mu  (2 m_{a}^2+3 m_{b}^2) }{2 m_{a}^4 m_{b}^3}-g_{\mathrm{P}}^{a} g_{\mathrm{S}}^{b}\frac{\mu  (3 m_{a}^2+2 m_{b}^2)  }{2 m_{a}^3 m_{b}^4},\\
h_{13,6}^{\textrm{NR}}=&g_{\mathrm{S}}^{a}g_{\mathrm{P}}^{b} \frac{69  \mu   (m_{a}^2+m_{b}^2) }{8 m_{a}^4 m_{b}^3 r}-g_{\mathrm{P}}^{a} g_{\mathrm{S}}^{b}\frac{69 \mu   (m_{a}^2+m_{b}^2) }{8 m_{a}^3 m_{b}^4 r}.
\end{aligned}
\end{equation}
The coefficients $h_{14,j}^{\textrm{NR}}$ are given by
\begin{equation}
\begin{aligned}
h_{14,1}^{\textrm{NR}}=&- g_{\mathrm{S}}^{a}g_{\mathrm{P}}^{b}  \frac{\mu ^2[m_{a}^2 (m_0^3 r^3-2 m_0^2 r^2-9 m_0 r-9)-3 m_{b}^2 C_2] }{32 m_{a}^4 m_{b}^3 r^4}\\&- g_{\mathrm{P}}^{a} g_{\mathrm{S}}^{b}\frac{ \mu ^2[3 m_{a}^2 C_2+m_{b}^2 (-m_0^3 r^3+2 m_0^2 r^2+9 m_0 r+9)] }{32 m_{a}^3 m_{b}^4 r^4},\\
h_{14,2}^{\textrm{NR}}=&0,\\
h_{14,3}^{\textrm{NR}}=& g_{\mathrm{S}}^{a}g_{\mathrm{P}}^{b}\frac{ \mu ^2 C_2}{16 m_{a}^2 m_{b}^3 r^3}-g_{\mathrm{P}}^{a} g_{\mathrm{S}}^{b}\frac{ \mu ^2 C_2}{16 m_{a}^3 m_{b}^2 r^3},\\
h_{14,4}=&- g_{\mathrm{S}}^{a}g_{\mathrm{P}}^{b}\frac{3  \mu ^2(3 m_{a}^2+4 m_{b}^2)  }{8 m_{a}^4 m_{b}^3 r}+g_{\mathrm{P}}^{a} g_{\mathrm{S}}^{b} \frac{3  \mu ^2 (4 m_{a}^2+3 m_{b}^2) }{8 m_{a}^3 m_{b}^4 r},\\
h_{14,5}^{\textrm{NR}}=&0,\\
h_{14,6}^{\textrm{NR}}=& -g_{\mathrm{S}}^{a}g_{\mathrm{P}}^{b} \frac{  \mu ^2  }{m_{a}^2 m_{b}^3}+g_{\mathrm{P}}^{a} g_{\mathrm{S}}^{b} \frac{\mu ^2  }{m_{a}^3 m_{b}^2}.
\end{aligned}
\end{equation}
The coefficients $h_{15,j}^{\textrm{NR}}$ are given by
\begin{equation}
\begin{aligned}
h_{15,1}^{\textrm{NR}}=&0,\\
h_{15,2}^{\textrm{NR}}=&0,\\
h_{15,3}^{\textrm{NR}}=& g_{\mathrm{P}}^{a} g_{\mathrm{S}}^{b}\frac{ C_2}{8 m_{a} m_{b}^2 r^3},\\
h_{15,4}^{\textrm{NR}}=&0,\\
h_{15,5}^{\textrm{NR}}=&0,\\
h_{15,6}^{\textrm{NR}}=&- g_{\mathrm{P}}^{a} g_{\mathrm{S}}^{b}\frac{2  }{m_{a} m_{b}^2}.
\end{aligned}
\end{equation}
The coefficients $h_{16,j}^{\textrm{NR}}$ are given by
\begin{equation}
\begin{aligned}
h_{16,1}^{\textrm{NR}}=&-g_{\mathrm{S}}^{a}g_{\mathrm{P}}^{b} \frac{ \mu [3 m_0^2(m_{a}^2+m_{b}^2) +8 m_{a}^2 m_{b}^2]  C_1}{64 m_{a}^4 m_{b}^3 r^3}\\&+g_{\mathrm{P}}^{a} g_{\mathrm{S}}^{b}\frac{ \mu  [3m_0^3r  (m_{a}^2+m_{b}^2) +m_0^2(8 m_{b}^2m_{a}^2 r^2+3m_{a}^2 +3 m_{b}^2) +16  m_0m_{a}^2 m_{b}^2 r+16 m_{a}^2 m_{b}^2]}{64 m_{a}^3 m_{b}^4 r^3},\\
h_{16,2}^{\textrm{NR}}=&g_{\mathrm{S}}^{a}g_{\mathrm{P}}^{b} \frac{ \mu(2 m_{a}^2+3 m_{b}^2)   C_1}{32 m_{a}^4 m_{b}^3 r^3}-g_{\mathrm{P}}^{a} g_{\mathrm{S}}^{b}\frac{ \mu (3 m_{a}^2+2 m_{b}^2)  C_1}{32 m_{a}^3 m_{b}^4 r^3},\\
h_{16,3}^{\textrm{NR}}=&g_{\mathrm{S}}^{a}g_{\mathrm{P}}^{b} \frac{3   \mu(m_{a}^2+m_{b}^2)  C_2}{32 m_{a}^4 m_{b}^3 r^4}-g_{\mathrm{P}}^{a} g_{\mathrm{S}}^{b}\frac{3  \mu (m_{a}^2+m_{b}^2)  C_2}{32 m_{a}^3 m_{b}^4 r^4},\\
h_{16,4}^{\textrm{NR}}=&g_{\mathrm{S}}^{a}g_{\mathrm{P}}^{b} \frac{ \mu [ m_{a}^2(3 m_0^2 r^2+8 m_{b}^2 r^2+9)+3 m_{b}^2 (m_0^2 r^2+3)]}{16 m_{a}^4 m_{b}^3 r^2}\\
&-g_{\mathrm{P}}^{a} g_{\mathrm{S}}^{b} \frac{3 \mu  [m_{a}^2(m_0^2 r^2+8 m_{b}^2 r^2+3) +m_{b}^2 (m_0^2 r^2+3)]}{16 m_{a}^3 m_{b}^4 r^2},\\
h_{16,5}^{\textrm{NR}}=&-g_{\mathrm{S}}^{a}g_{\mathrm{P}}^{b} \frac{ \mu  (2 m_{a}^2+3 m_{b}^2) }{8 m_{a}^4 m_{b}^3}+g_{\mathrm{P}}^{a} g_{\mathrm{S}}^{b}\frac{  \mu   (3 m_{a}^2+2 m_{b}^2)}{8 m_{a}^3 m_{b}^4},\\
h_{16,6}^{\textrm{NR}}=&- g_{\mathrm{S}}^{a}g_{\mathrm{P}}^{b} \frac{3\mu  (m_{a}^2+m_{b}^2)  }{2 m_{a}^4 m_{b}^3 r}+ g_{\mathrm{P}}^{a} g_{\mathrm{S}}^{b}\frac{3  \mu  (m_{a}^2+m_{b}^2)}{2 m_{a}^3 m_{b}^4 r}.
\end{aligned}
\end{equation}

\subsection{Useful formulas in Fourier Transformation}\label{sec.fourier}
\setcounter{equation}{0}
\renewcommand\theequation{B.\arabic{equation}}
The potential in coordinate space is defined as
\begin{equation}
V[\boldsymbol{r},-i\frac{\partial}{\partial\boldsymbol{\bar{r}}}]\equiv\mathcal{F}[\overline{V}(\boldsymbol{q},\boldsymbol{p}_1)] \equiv \Big[\int \frac{d^3\boldsymbol{q}}{(2\pi)^3}e^{i\boldsymbol{q}\cdot \boldsymbol{\bar{r}}}\overline{V}(\boldsymbol{q},\boldsymbol{p}_1)\Big]\Big|_{\textrm{place $\boldsymbol{p}_1$ at far right and replace $\boldsymbol{p}_1$ with $-i\frac{\partial}{\partial\boldsymbol{\bar{r}}}$}}.
\end{equation}
Some useful formulas for the Fourier transformation are:
\begin{equation}
\begin{aligned}
\mathcal{F}[\frac{1}{\boldsymbol{q}^2+m_{0}^2}]=& \frac{1}{4\pi r}e^{-m_{0}r}\equiv W\frac{1}{r},\\
\mathcal{F}[\frac{\boldsymbol{q}_i}{\boldsymbol{q}^2+m_{0}^2}]=&iWr^2C_1\boldsymbol{\bar{r}}_i,\\
\mathcal{F}[\frac{\boldsymbol{q}_i\boldsymbol{q}_j}{\boldsymbol{q}^2+m_{0}^2}]=& Wr^3[C_{21}T_{ij,1}+C_{22}T_{ij,2}],\\
\mathcal{F}[\frac{\boldsymbol{q}_i\boldsymbol{q}_j\boldsymbol{q}_k}{\boldsymbol{q}^2+m_{0}^2}]=& iWr^4[C_{31}T_{ijk,1}+C_{32}T_{ijk,2}],\\
\mathcal{F}[\frac{\boldsymbol{q}_i\boldsymbol{q}_j\boldsymbol{q}_k\boldsymbol{q}_l}{\boldsymbol{q}^2+m_{0}^2}]=& Wr^5[C_{41}T_{ijkl,1}+C_{42}T_{ijkl,2}+C_{43}T_{ijkl,3}],
\end{aligned}
\label{eq:Fourier-transformation}
\end{equation}
where $W=\frac{1}{4\pi}e^{-m_{0}r}$, and the coefficients are expressed as
\begin{equation}
\begin{aligned}
C_{21} =&-C_2+16\pi r^3\delta^3(\boldsymbol{r}),\\
C_{22} =&C_1-4\pi r^3\delta^3(\boldsymbol{r}),\\
C_{31} =&-[m_{0}^3r^3+6m_{0}^2r^2+15m_{0}r+15r]+92\pi r^3\delta^3(\boldsymbol{r}),\\
C_{32} =&-C_{21},\\
C_{41} =&[m_{0}^4r^4+10m_{0}^3r^3+45m_{0}^2r^2+105m_{0}r+105]-[44(m_{0}^2r^2+1)+704]\pi r^3\delta^3(\boldsymbol{r}),\\
C_{42} =&C_{31}+4\pi m_0^2r^5\delta^3(\boldsymbol{r}),\\
C_{43} =&-C_{21},
\end{aligned}
\end{equation}
with $C_{1,2}$ defined in Eq.~(\ref{eq:V0C1C2}).
The tensor structures in Eq.~\eqref{eq:Fourier-transformation} are defined as
\begin{equation}
\begin{aligned}
T_{ij,1}&\equiv\boldsymbol{\bar{r}}_i\boldsymbol{\bar{r}}_j,\\
T_{ij,2}&\equiv\delta_{ij},\\
T_{ijk,1}&\equiv\boldsymbol{\bar{r}}_i\boldsymbol{\bar{r}}_j\boldsymbol{\bar{r}}_k,\\
T_{ijk,2}&\equiv\boldsymbol{\bar{r}}_i\delta_{jk}+\boldsymbol{\bar{r}}_j\delta_{ik}+\boldsymbol{\bar{r}}_k\delta_{ij},\\
T_{ijkl,1}&\equiv\boldsymbol{\bar{r}}_i\boldsymbol{\bar{r}}_j\boldsymbol{\bar{r}}_k\boldsymbol{\bar{r}}_l,\\
T_{ijkl,2}&\equiv\boldsymbol{\bar{r}}_i\boldsymbol{\bar{r}}_j\delta_{kl}+
\boldsymbol{\bar{r}}_i\boldsymbol{\bar{r}}_k\delta_{jl}+
\boldsymbol{\bar{r}}_i\boldsymbol{\bar{r}}_l\delta_{jk}+
\boldsymbol{\bar{r}}_j\boldsymbol{\bar{r}}_k\delta_{il}+
\boldsymbol{\bar{r}}_j\boldsymbol{\bar{r}}_l\delta_{ik}+
\boldsymbol{\bar{r}}_k\boldsymbol{\bar{r}}_l\delta_{ij}, \\
T_{ijkl,3}&\equiv\delta_{ij}\delta_{kl}+\delta_{ik}\delta_{kl}+\delta_{il}\delta_{jl}.
\end{aligned}
\end{equation}
%The Dirac $\delta$ function is understood as
%\begin{equation}
%\begin{aligned}
%\int_{-\infty}^{\infty}f(\boldsymbol{r})\delta^3({\boldsymbol{r}})d^3\boldsymbol{r}&=2f(\boldsymbol{0})\int_0^{\infty} \delta(r)dr=f(\boldsymbol{0}),\\
%\delta(\boldsymbol{r}) &=\frac{1}{2\pi r^2}\delta(r)
%\end{aligned}
%\end{equation}
%\com{XXXXXX  To directly replace the notation of $\delta^3({\boldsymbol{r}})$ with $\frac{1}{2\pi r^2}\delta(r)$ in all the formulas? XXXXXX}

Furthermore, the following relations hold:
\begin{equation}
\begin{aligned}
\int\delta^3(\boldsymbol{r})T_{ij,1}d\Omega =&\frac{1}{3}\int\delta^3(\boldsymbol{r})T_{ij,2}d\Omega,\\
\int\delta^3(\boldsymbol{r})T_{ijk,\alpha}d\Omega =&\int\delta^3(\boldsymbol{r})T_{ijk,1\alpha}Y_{1m}Y^{*}_{1m'}d\Omega=0,\\
\int\delta^3(\boldsymbol{r})T_{ijkl,1}d\Omega =&\frac{1}{15}\int\delta^3(\boldsymbol{r})T_{ijkl,3}d\Omega,\\
\int\delta^3(\boldsymbol{r})T_{ijkl,2}d\Omega =&\frac{2}{3}\int\delta^3(\boldsymbol{r})T_{ijkl,3}d\Omega,
\end{aligned}
\label{eq:propety-of-delta-function-2}
\end{equation}
with $\alpha=1,2$.

For the atomic systems, the properties of the radial components of the wave functions impose specific constraints on the contributions of various terms. Specifically, $\delta^3(\boldsymbol{r})T_{ij,\alpha}$ contributes exclusively to $S$-wave states, $\delta^3(\boldsymbol{r})T_{ijk,\alpha}$ contributes to $S$-wave and $P$-wave states, and $\delta^3(\boldsymbol{r})T_{ijkl,\alpha}$ contributes to $S$-, $P$-, and $D$-wave states. Consequently, one can express:
\begin{equation}
\begin{aligned}
\delta^3(\boldsymbol{r})T_{ij,1}=& \frac{1}{3}\delta^3(\boldsymbol{r})T_{ij,2},\\
\delta^3(\boldsymbol{r})T_{ijk,\alpha}=& 0.
\end{aligned}
\end{equation}

\bibliography{axion-potential}
\bibliographystyle{apsrev4-2}

\end{document}